\documentclass{article}

\usepackage[english]{babel}
\usepackage{subcaption}
\usepackage{caption}
\usepackage{multirow}
\usepackage{svg}
\usepackage{booktabs} 
\usepackage{array}
\usepackage{amsfonts} 
\usepackage{amssymb}  
\usepackage{booktabs} 
\usepackage[numbers,sort&compress]{natbib}

\usepackage[letterpaper,top=2cm,bottom=2cm,left=3cm,right=3cm,marginparwidth=1.75cm]{geometry}

\usepackage{amsmath}
\usepackage{graphicx}
\usepackage[colorlinks=true, allcolors=blue]{hyperref}

\title{EUR/USD Exchange Rate Forecasting Based on Information Fusion with Large Language Models and Deep Learning Methods}
\author{Hongcheng Ding*, Xuanze Zhao, Ruiting Deng, Shamsul Nahar Abdullah, \\
Deshinta Arrova Dewi\\
INTI International University, Malaysia\\
i24025877@student.newinti.edu.my}

\begin{document}
\maketitle

\begin{abstract}
Accurate forecasting of the EUR/USD exchange rate is crucial for investors, businesses, and policymakers. This paper proposes a novel framework, IUS, that integrates unstructured textual data from news and analysis with structured data on exchange rates and financial indicators to enhance exchange rate prediction. The IUS framework employs large language models for sentiment polarity scoring and exchange rate movement classification of texts. These textual features are combined with quantitative features and input into a Causality-Driven Feature Generator. An Optuna-optimized Bi-LSTM model is then used to forecast the EUR/USD exchange rate. Experiments demonstrate that the proposed method outperforms benchmark models, reducing MAE by 10.69\% and RMSE by 9.56\% compared to the best performing baseline. Results also show the benefits of data fusion, with the combination of unstructured and structured data yielding higher accuracy than structured data alone. Furthermore, feature selection using the top 12 important quantitative features combined with the textual features proves most effective. The proposed IUS framework and Optuna-Bi-LSTM model provide a powerful new approach for exchange rate forecasting through multi-source data integration.


\end{abstract}

\section{Introduction}

The exchange rate between the Euro and the US Dollar is a significant indicator in the global financial market, reflecting the economic dynamics between two of the world's largest economies. Precise prediction of the EUR/USD exchange rate is crucial for individual investors, businesses engaged in international trade, and policymakers responsible for economic stability and growth. Traditionally, econometric models have been utilized to forecast exchange rates, relying heavily on historical market data and macroeconomic indicators released by governments and financial organizations \cite{rossi2013exchange}. Although these datasets are comprehensive, their low publication frequency makes it difficult to capture real-time market volatility and nonlinear dynamics \cite{cheung2019exchange}.

The integration of unstructured data from diverse sources, such as news articles, financial reports and social media platforms, has the potential to improve the accuracy of exchange rate forecasting. In recent years, the significant impact of political events, global economic crises, and unexpected international incidents on currency fluctuations has been recognized \cite{sezer2020financial}, suggesting that considering a wider range of information beyond traditional structured data may be beneficial. The great amount of textual data may contain valuable insights into market sentiment, economic trends, and key events that can influence exchange rates \cite{hu2021survey}. However, exchange rate forecasting presents two significant challenges. Firstly, while the relationship between news information and market trends is relatively straightforward in traditional financial markets, the complex semantics of market-driven news and analysis texts in the context of exchange rates pose difficulties for sentiment analysis. Secondly, traditional methods struggle to adequately capture the complex nonlinear patterns and unstructured relationships hidden within the textual data, limiting their ability to provide accurate predictions in the dynamically changing foreign exchange market\cite{semiromi2020news}.

Recent attempts to address these challenges by integrating textual data and deep learning methods show promise, but limitations still exist. Singh et al. \cite{singh2024utilising} propose using Word2Vec and LSTM to classify collected Weibo text data and assign sentiment weights to each word. The sentiment analysis results are then incorporated into a CNN-LSTM hybrid model for exchange rate prediction. However, the texts related to exchange rates often contain news or comments about two countries simultaneously, making it challenging to accurately attribute the content to a specific country. The existence of massive noise and the involvement of information related to two currencies in the text used for exchange rate forecasting create challenges in handling lengthy texts and complex semantics. Tadphale et al. \cite{tadphale2023impact} utilize news headlines for sentiment analysis and then combine with other market indicators as inputs to an LSTM model for exchange rate prediction. While this approach shows potential, news headlines have limited information content and lack complete context due to text length. Moreover, the news of each day may also have a potential impact on the exchange rate movement of the next day, which is not fully captured by the proposed approach.

In this paper, we present a novel framework that integrates unstructured and structured (IUS) data to forecast EUR/USD exchange rates. We use ChatGPT-4.0 to filter out noise from collected news and analysis texts, extracting segments related to the exchange rate, in order to obtain the initial dataset. The dataset is then annotated with sentiment polarity scores by ChatGPT-4.0 and next-day exchange rate movements, e.g. up or down. Subsequently, we fine-tune two large language models (LLMs) on this dataset to create textual features, textual sentiment and exchange rate movement features. Next, the collected exchange rate and financial market data are utilized to generate quantitative features, which are integrated with textual features and inputted into a Causality-Driven Feature Generator. Finally, all generated features are fed into an Optuna-Bi-LSTM model to predict the EUR/USD exchange rate.

Our proposed method effectively addresses the challenges of processing news articles containing information about both countries in an exchange rate pair, handling high noise, complex semantics, the lack of contextual information in brief texts, and the various extraction of textual features. We validate the effectiveness of our proposed exchange rate prediction approach on our EUR/USD exchange rate dataset. The results demonstrate that our model outperforms the strongest benchmark models by 10.69\% in terms of Mean Absolute Error (MAE) and by 9.56\% in terms of Root Mean Squared Error (RMSE). As for data fusion, by combining unstructured and structured data, the Optuna-Bi-LSTM model is able to enhance prediction accuracy beyond what is possible with structured data alone. Furthermore, using the top 12 important features selected by the RFE method combined with 31 textual features proves to be more effective compared to analyzing all textual features, as it more directly corresponds to the actual exchange rate response to market conditions.

This research makes several significant contributions to the field of exchange rate prediction:
\begin{itemize}
\itemsep=0pt
  \item We propose a novel IUS framework that integrates unstructured and structured data to predict exchange rates, setting a new benchmark in the industry.
  \item By utilizing LLMs, we develop an annotated EUR/USD exchange rate text dataset through a multi-step processing of raw textual data, introducing a new approach to sentiment analysis for complex semantic texts.
  \item Through extensive literature review and the application of a Causality-Driven Feature Generator, we construct a comprehensive feature set that incorporates a wide range of economic, financial, and textual indicators.
  \item To capture the nonlinear dynamics of exchange rate time series, we employ a Bi-ISTM deep learning model which is further optimized by the Optuna hyperparameter optimization framework. This ensures optimal predictive performance and enhances the generalizability of our methods.
\end{itemize}
\section{Motivation}
\begin{figure}[ht]
\centering
\includegraphics[width=\linewidth, keepaspectratio]{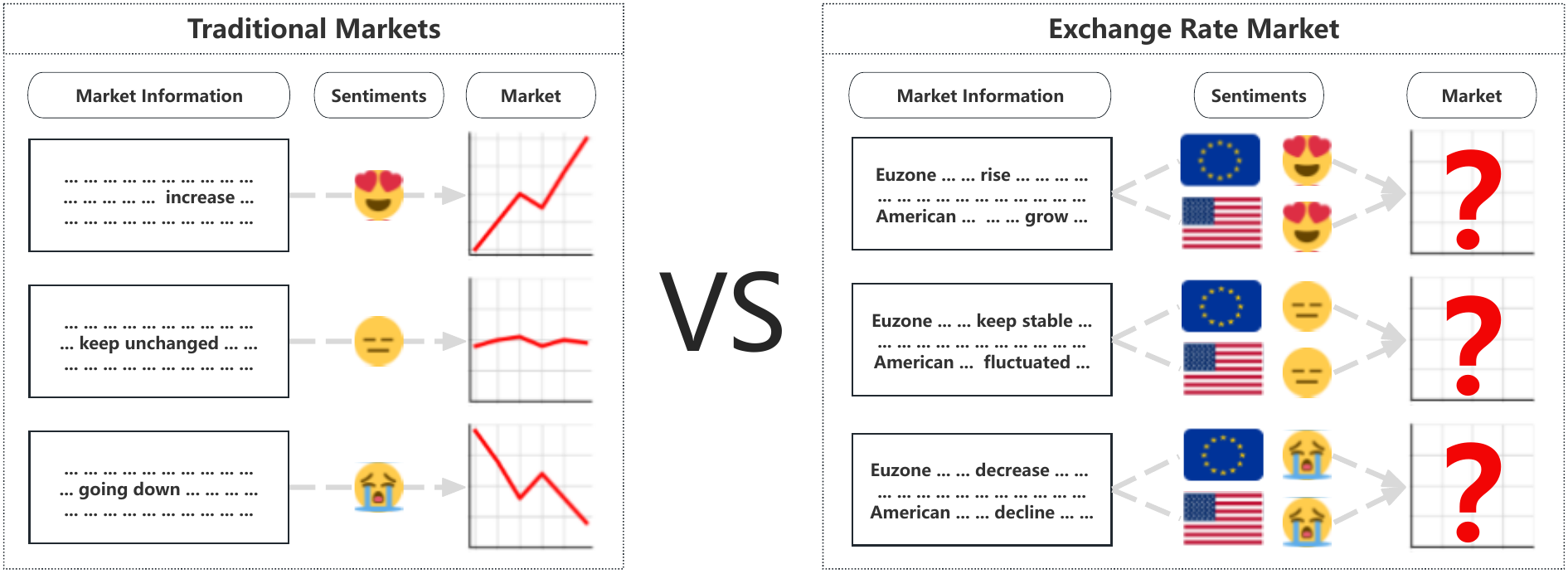} 
\caption{\label{fig:constrast}Comparative analysis of sentiment impact on traditional vs. exchange rate markets.}
\end{figure}
Traditional sentiment analysis methods in exchange rate prediction have primarily focused on brief texts such as Twitter posts and news headlines. These methods neglect the rich semantics and diverse information contained within longer content, which are essential for a thorough understanding of market sentiment. Additionally, the effectiveness of dictionary-based sentiment analysis heavily depends on the quality and comprehensiveness of predefined word lists. These methods face challenges in adapting to a changing market because sentiment dictionaries tend to be static and are not immediately updated, thus failing to reflect domain-specific vocabulary. More critically, traditional methods struggle to capture the differences in numerical data and its implications for market sentiment. However, in a financial context, the extent of data growth and the variations in percentages often reflect market fluctuations. For instance, a 0.005\% increase in a specific data point might influence the market differently compared to a 50\% increase.

The relationship between news information and market trends is relatively straightforward in traditional markets, as illustrated in Figure \ref{fig:constrast}. However, the complex semantics of news and analysis texts in the context of exchange rates pose difficulties for sentiment analysis. The nature of financial texts makes sentiment analysis and annotation a particularly challenging task, as it often contains specialized jargon, implicit sentiments linked to market conditions, and subtle variations across different sectors. Moreover, exchange rates involve two countries, and the impact of primarily positive or negative news on financial markets can be significant. For instance, an abundance of positive news about the U.S. can lead to a strengthening of the U.S. dollar, which typically results in a decrease in the EUR/USD exchange rate, and vice versa. However, news texts often contain associated information related to both countries, making classification difficult, and the positive or negative news from both countries has a zero-sum relationship in the final sentiment analysis.
\begin{figure}[ht]
\centering
\includegraphics[width=\linewidth, keepaspectratio]{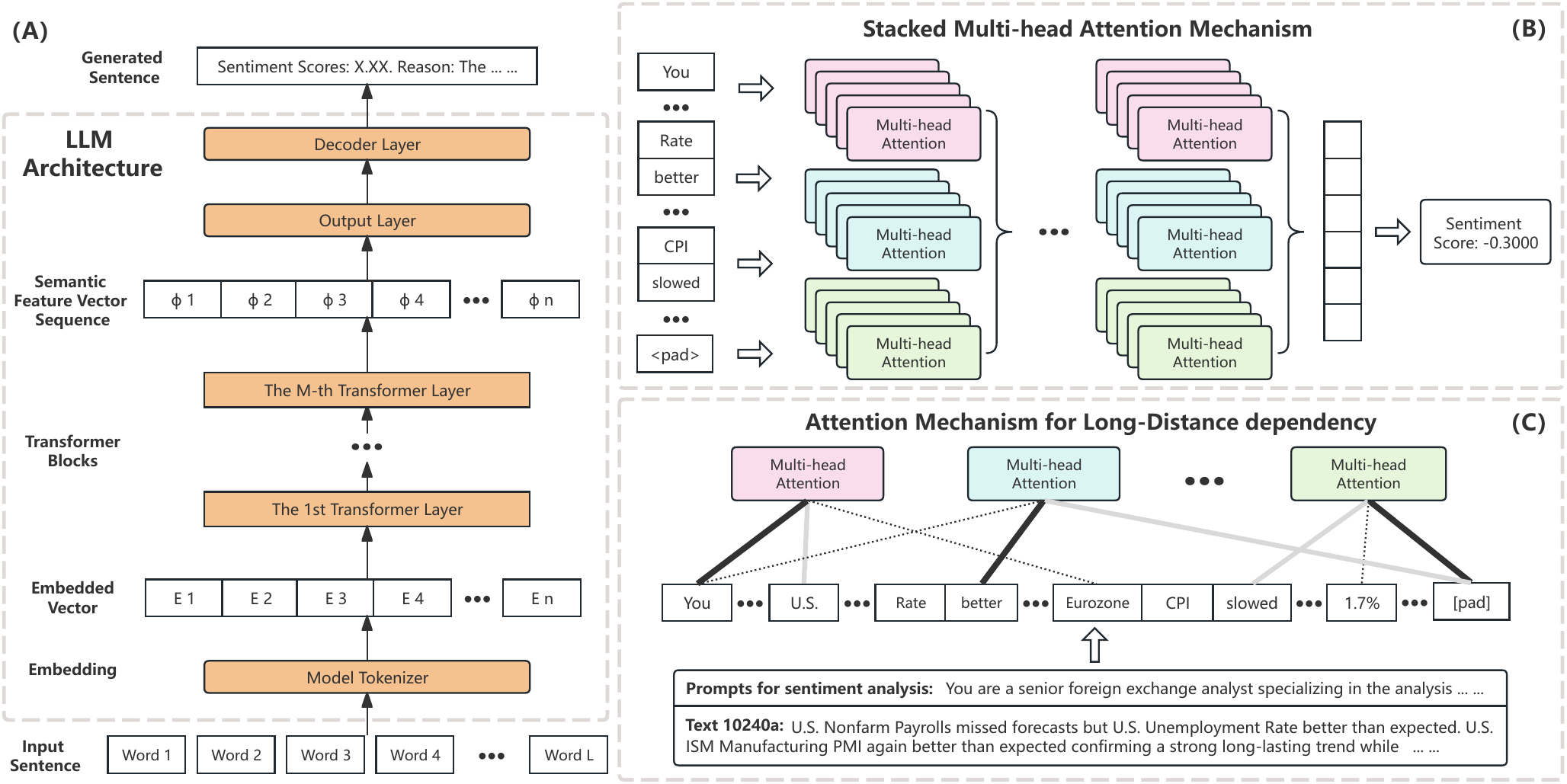} 
\caption{\label{fig:Long-distance}Capturing long-distance dependencies in large language models (LLMs): (A) LLM architecture with stacked transformer blocks and multi-head attention mechanisms; (B) Attention mechanism focusing on different positions within the input sequence; (C) Stacked multi-head attention mechanism maintaining focus on relevant elements throughout the lengthy text.}
\end{figure}

To tackle the challenges of sentiment analysis in extensive financial texts, we utilize the capabilities of LLMs. As depicted in Figure \ref{fig:Long-distance}(A), LLMs are structured with deep and hierarchical Transformer blocks, each featuring a multi-head self-attention mechanism. This architecture allows LLMs to recognize complex patterns and capture long-range dependencies within the text. As texts progress through these Transformer blocks, illustrated in Figure \ref{fig:Long-distance}(B), the multiple attention heads within a block enable simultaneous focus on various textual elements. The different lines represent the varying attention paid by the multi-head attention mechanism to elements at different positions, revealing complex interrelationships even among distant sentences. Following this, the layered structure shown in Figure \ref{fig:Long-distance}(C), with its consistent application of self-attention, effectively addresses the complexities of lengthy financial texts. By recognizing long-distance dependencies and maintaining focus on relevant elements throughout the input, LLMs can analyze every detail of the text to generate an overall sentiment polarity score. Moreover, LLMs have the ability to capture the differences in numerical data, enabling them to provide more accurate sentiment polarity scores that reflect the actual impact of these figures on market sentiment.
\begin{figure}[h]
\centering
\includegraphics[width=\linewidth, keepaspectratio]{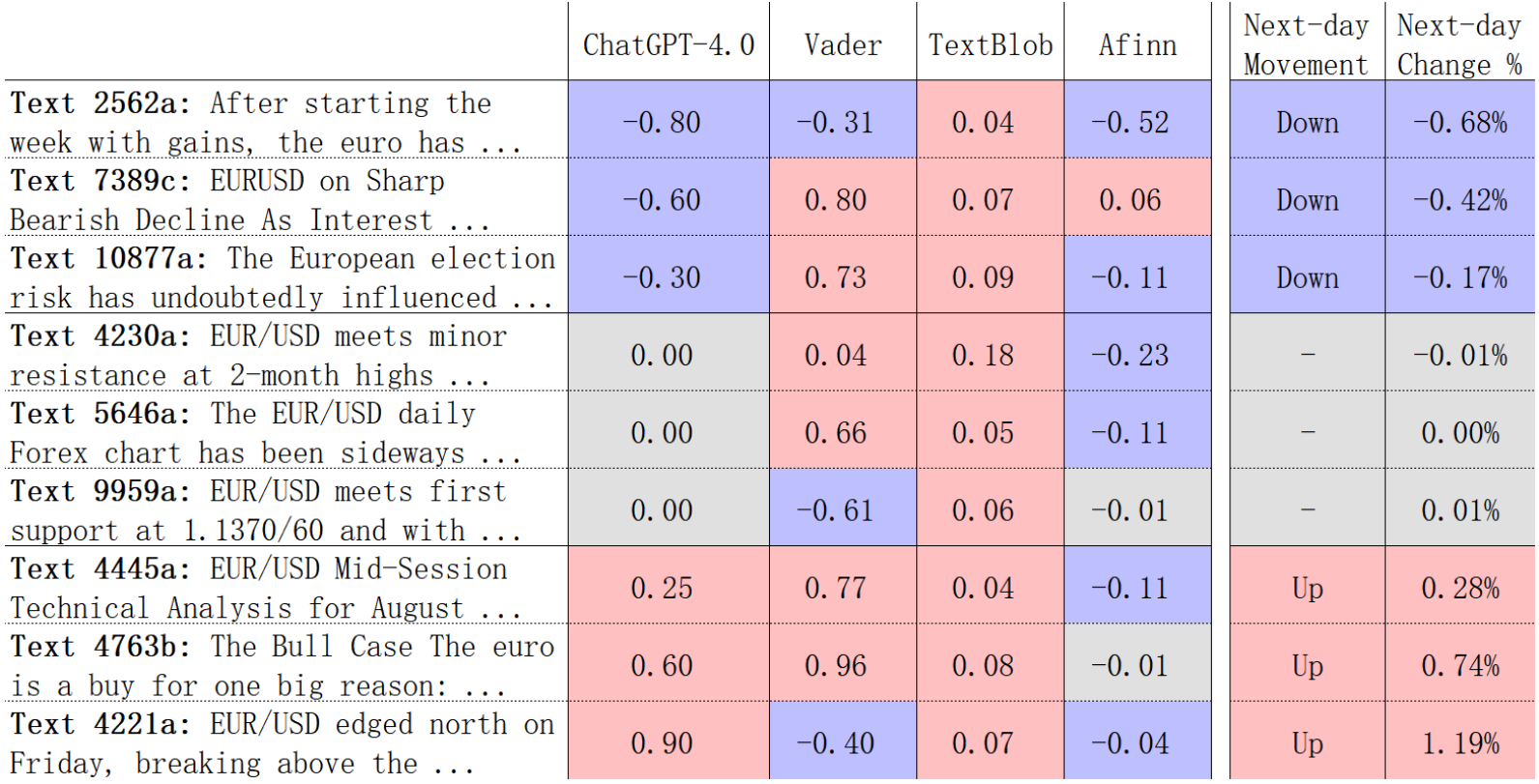} 
\caption{\label{fig:comparison1} Comparative analysis of sentiment scores by ChatGPT-4.0 and traditional tools.}
\end{figure}

We employ ChatGPT-4.0 and several traditional dictionary-based sentiment analysis tools, including Vader, TextBlob, and Afinn, to conduct sentiment analysis on collected news and analysis texts and compare these data with trends in the forex market. As demonstrated in Figure \ref{fig:comparison1}, the traditional tools exhibit poor performance when processing longer texts, exhibiting significant variety in the sentiment scores they produce, which correlate poorly with actual market movements in forex trading. In contrast, the results from ChatGPT-4.0 show a higher consistency with both the next-day movements and change percentages.
\section{Related Work}

1.Current studies on exchange rate forecasting

Haider et al. \cite{haider2023commodity} examine whether commodity prices can forecast exchange rates in commodity-dependent economies using both in-sample and out-of-sample techniques. By modeling commodity prices to predict USD rates, their findings indicate this approach is more effective than a random walk model, providing valuable perspectives across various economies. Sarkar and Ali \cite{sarkar2022eur} analyze linear regression for predicting EUR/USD exchange rates using normalized daily and hourly data. Their research applies this approach to different time series, offering strategies to help traders mitigate issues and enhance profitability in the forex market. Ruan et al. \cite{ruan2024forecasting} evaluate whether economic policy uncertainty (EPU) outperforms traditional macroeconomic indicators in predicting exchange rate volatility in both developed and emerging markets. Their results demonstrate the superior predictive capability of EPU, suggesting significant implications for risk management and policy-making, and recommending broader application to verify these findings' generalizability. Windsor and Cao \cite{windsor2022improving} develop a comprehensive system using market indicators and investor sentiments to predict the USD/CNY exchange rate. This innovative system effectively captures complex interactions among various financial factors, providing a precise and robust forecasting tool. Salisu et al. \cite{salisu2021oil}  demonstrate that oil prices are a reliable indicator of exchange rate returns for both net oil exporters and importers. Their study emphasizes the importance of considering asymmetries in the data, which substantially enhances the predictability of an oil-based model. The results underscores the potential of oil prices as a crucial factor in financial forecasting models. Neghab et al. \cite{neghab2024explaining} employ machine learning techniques, including linear regression, tree-based models, and deep learning, to forecast exchange rates based on macroeconomic fundamentals. The study addresses challenges such as nonlinearity, multicollinearity, time variation, and noise in modeling.

2.The application of unstructured data in predictions

Ito and Takeda \cite{ito2022sentiment} improve the accuracy of exchange rate models by using sentiment indices constructed from Google search volumes of financial terms. This approach effectively captures timely market sentiments, although the generalizability of their findings requires further exploration due to the analysis's limited scope. Ben Omrane et al. \cite{ben2020dynamic} investigate the impact of US and EU macroeconomic news on the volatility and returns of the EUR/USD exchange rate using regime smooth transition regression. Their findings indicate that the effects of news vary between economic states, with US news generally having a larger impact than EU news on currency fluctuations. Li et al. \cite{li2019text} introduce a novel approach for forecasting crude oil prices that incorporates online news text mining to extract sentiment features and group news by topic, effectively capturing the immediate market impacts of various events. This method enriches traditional forecasting models, although it is limited by its reliance on a single news source and potential noise in online news data. Bai et al. \cite{bai2022crude} propose a robust framework for forecasting crude oil prices that exploits advanced text mining techniques on news headlines to construct high-quality features. They introduce novel indicators for topic and sentiment analysis tailored for short texts, enhancing the model's performance. However, the study does not deeply investigate the relative importance of textual features compared to non-textual factors. Swathi et al. \cite{swathi2022optimal} employ Twitter sentiment analysis to predict stock prices by analyzing emotions and opinions in stock-related tweets. This method provides deeper insights into market sentiment and enhances the predictive performance of their model, although it predominantly relies on Twitter, suggesting the potential benefits of incorporating more diverse data sources. Kalamara et al. \cite{kalamara2022making} explore the use of newspaper text to extract economic signals, demonstrating that such information can substantially improve macroeconomic forecasts. By combining a large array of text-derived regressors with supervised machine learning, they achieve significant forecast improvements, especially during periods of economic stress. Naeem et al. \cite{naeem2021machine} utilize a machine learning approach to forecast the USD/PKR exchange rate, employing sentiment analysis of finance-related tweets and various machine learning classifiers to process and optimize the dataset. This innovative method showcases the potential of leveraging social media data for financial predictions. Lv et al. \cite{xueling2023exchange} develop a hybrid model that combines sentiment analysis of Weibo text data with historical exchange rate information to predict market trends. This method improves prediction accuracy by integrating the perspectives of market participants, demonstrating the substantial impact of social media sentiment on forecasting exchange rates. Küçüklerli and Ulusoy \cite{kuccuklerli2024sentiment} integrate Twitter sentiment analysis with economic indicators to predict the USD/TL exchange rate using machine learning techniques. They collect exchange rate data and finance-related tweets, preprocess the data, and employ various ML models. The LSTM model achieves the highest accuracy of 65\% in forecasting the exchange rate. Semiromi et al. (2024) predict currency pairs using news story events from the economic calendar and machine learning techniques. They use text mining methods, sentiment analysis with a new sentiment dictionary, and machine learning algorithms, achieving over 60\% accuracy in predicting exchange rate movements.

3. Traditional Predictive Methods

Traditional econometric models and machine learning techniques have been extensively applied to exchange rate forecasting. Li et al. \cite{li2023analysis} utilize ARIMA and GARCH models to predict USD/EUR exchange rate fluctuations, showing GARCH's effectiveness in capturing financial data volatility while noting the limitations of ARIMA in exchange rate forecasting. Zhang \cite{zhang2024rmb} applies Simple Exponential Smoothing and ARIMA models to forecast the RMB/USD exchange rate, exploring their advantages in capturing exchange rate dynamics and providing accurate predictions. Colombo and Pelagatti \cite{colombo2020statistical} use various statistical learning methods, including regularized regression splines, RF, and SVM, to assess exchange rate models' predictive power. The sticky price monetary model with error correction specification shows strong forecasting performance at different horizons, outperforming the random walk benchmark, providing insights into the non-linear relationship between exchange rates and fundamentals. Pfahler \cite{pfahler2021exchange} employs artificial neural networks and gradient-boosted decision trees to forecast exchange rate movements using macroeconomic fundamentals from purchasing power parity, uncovered interest rate parity, and monetary model theories. These machine learning models outperform the random walk benchmark, especially when time dummies are included, highlighting the potential of complex interactions between time dummies and fundamentals in exchange rate forecasting. Khoa and Huynh \cite{khoa2022predicting} apply SVR under the uncovered interest rate parity framework to forecast the VND/USD exchange rate during the COVID-19 pandemic. Combining the framework with the robust SVR technique demonstrates superior performance compared to OLS regression and random walk models. However, these econometric models and traditional machine learning techniques face limitations in capturing the complex, non-linear dynamics and high-dimensional relationships inherent in exchange rate data.

4. Modern methods

Sun et al. \cite{sun2020new} introduce a methodology that combines LSTM with bagging ensemble learning, significantly enhancing the accuracy and profitability of exchange rate forecasts. This method outperforms traditional benchmark models, although it is solely applied to univariate exchange rate series without considering additional influencing factors. Liu et al. \cite{liu2024new} develop an innovative LASSO-Bi-LSTM ensemble learning strategy that merges the LASSO with Bi-LSTM networks for forecasting the USD/CNY exchange rate. This approach demonstrates superior accuracy over other deep learning models but is limited to short-term forecasts and is not tested in broader financial markets. Islam and Hossain \cite{islam2021foreign} craft a hybrid model that integrates GRU and LSTM networks to predict currency prices in the forex market for key pairs. The hybrid model surpasses the performance of LSTM, GRU and a simple moving average approach in terms of accuracy and risk-adjusted returns. Despite its success, the model sometimes struggles with abrupt price fluctuations and requires specific adjustments for optimal performance with the GBP/USD pair. Dautel et al. \cite{dautel2020forex} conduct an empirical analysis comparing various deep learning frameworks, including LSTM and GRU networks, for exchange rate prediction. The study provides valuable insights into the practical application of these models for financial forecasting, though it acknowledges challenges in model tuning and the disparity between statistical accuracy and economic relevancy. Wan et al. \cite{wan2019multivariate} introduce the Multivariate TCN, which uses a deep CNN with multichannel residual blocks and an asymmetric structure to enhance forecasting in aperiodic multivariate time series. This model shows marked improvements over other algorithms, and the study suggests that focusing on higher-order statistical features could simplify the model and boost performance. Zeng et al. \cite{zeng2023transformers} critically assess the efficacy of Transformer-based solutions for long-term time series forecasting, proposing a straightforward one-layer linear model, which outperforms more complex Transformer-based models across multiple datasets. However, the simplicity of this model limits its capacity. 
Optimization algorithms play a vital role in improving the performance of deep learning networks by systematically tuning their hyperparameters. Xu et al. \cite{xu2022research} propose a Particle Swarm Optimization with LSTM (PSO-LSTM) model that leverages particle swarm optimization to optimize LSTM hyperparameters, mimicking the social behavior of bird flocking to discover optimal solutions. Hamdia et al. \cite{hamdia2021efficient} employ a Genetic Algorithm, drawing inspiration from natural selection and evolution, to optimize deep neural network architectures. Victoria and Maragatham \cite{victoria2021automatic} utilize Bayesian Optimization, which constructs a probabilistic model of the objective function to guide the search for the best hyperparameters. Dong et al. \cite{dong2019dynamical} introduce an approach using Deep Reinforcement Learning, specifically a continuous Deep Q-learning algorithm with a heuristic strategy, for adaptive hyperparameter optimization in visual object tracking. García Amboage et al. \cite{amboage2024model} propose Swift-Hyperband, integrating performance prediction via SVR with early stopping methods to streamline the optimization process. Brodzicki et al. \cite{brodzicki2021whale} apply the Whale Optimization Algorithm (WOA), inspired by the foraging behavior of humpback whales, for deep neural network hyperparameter optimization.
\section{Methodology}
\subsection{The IUS Framework}
\begin{figure}[ht]
\centering
\includegraphics[width=\linewidth, keepaspectratio]{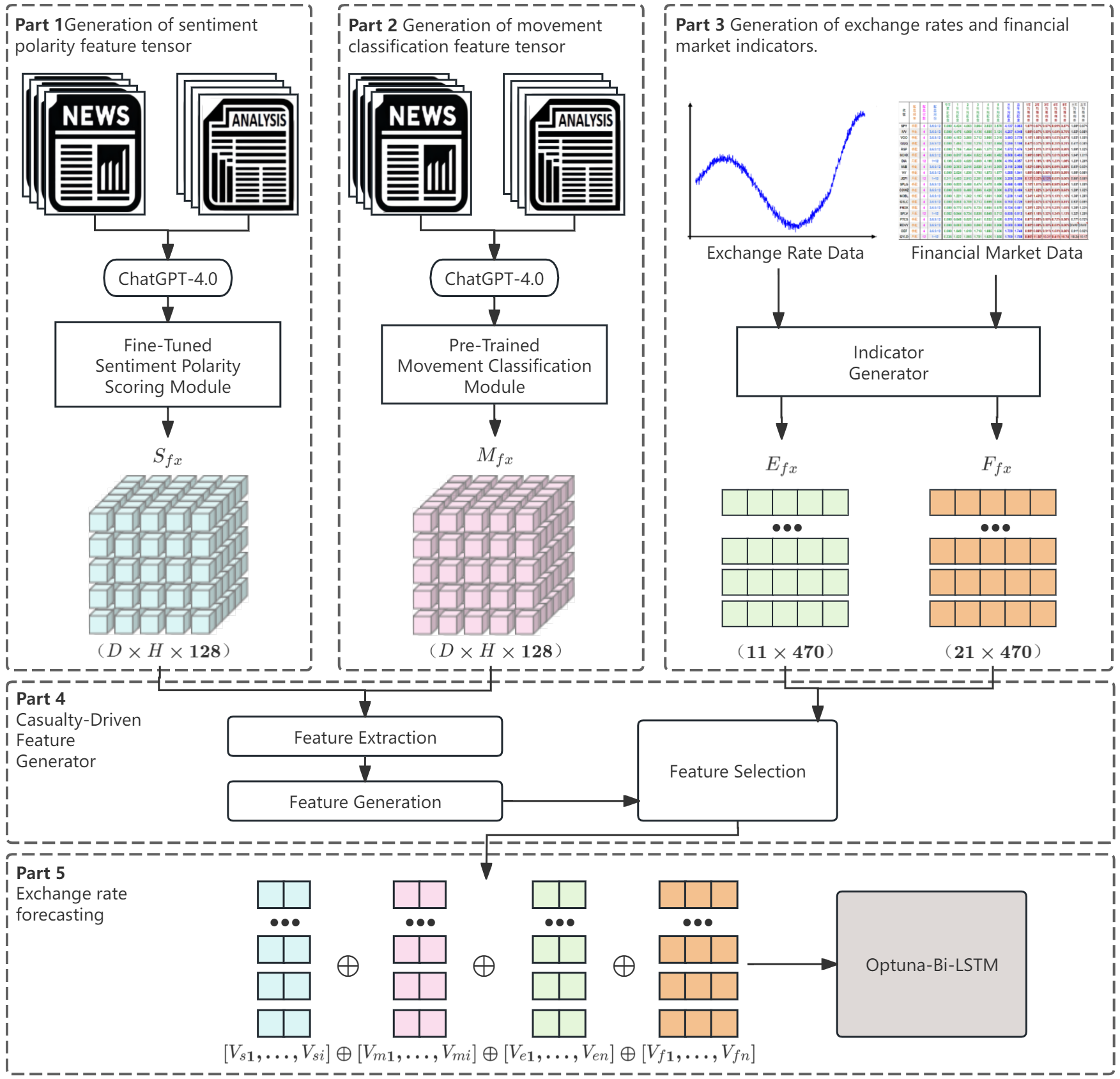} 
\caption{\label{fig:IUS} The framework of IUS to forecast EUR/USD exchange rate.}
\end{figure}

In this work, we introduce the IUS Framework, as illustrated in Figure \ref{fig:IUS}, which consists of five technical components. The first component is the Sentiment Polarity Scoring Module (SPSM), which employs an embedding generator based on a fine-tuned version of the RoBERTa-Large model, specifically adapted for sentiment analysis. This module generates the sentiment polarity feature tensor ($S_{fx}$) from news and analysis texts related to the target exchange rate over $D$ trading days, capturing the sentiment representation of all texts during this period. The second component, the Movement Classification Module (MCM), utilizes the original RoBERTa-Large model. Configured as an embedding generator, it produces the feature tensor ($M_{fx}$) for analyzing the next-day exchange rate movement, e.g., up or down, capturing the movement representation of texts within the same period. Subsequently, we extract quantitative indicator subsequences for the target exchange rate and its related exchange rates, as well as financial market indicators over $D$ trading days. $E_{fx}$ and $F_{fx}$ represent the quantitative features of these rates and financial indicators, respectively, over the same period. Finally, all these features are integrated and processed through a Causality-Driven Feature Generator, then input into the Bi-LSTM model to predict the closing price of the target exchange rate for the next trading day $y_{t+1}$.
\subsubsection{The IUS Framework}
For the textual dataset, we collect data from investing.com and forexempire.com, covering the period from February 6, 2016, to January 19, 2024. The dataset includes all accessible data on these platforms within the period, totaling 35,427 texts. However, due to the potential existence of noise and irrelevant information, we utilize ChatGPT-4.0 and prompt engineering techniques to filter the raw dataset. This data annotation approach aligns with prior research, such as using LLMs for automatic data annotation to detect hallucinations \cite{das2024compos}, keyword annotation and document content description generation \cite{gallipoli2024keyword}, and math problem knowledge tagging with few-shot learning \cite{li2024knowledge}. After filtering, we observe that the news and analysis texts inherently contain a higher level of noise, which may be attributed to the necessity of catering to the diverse needs of readers. In addition, we notice that often only individual paragraphs or multiple segments within an article are directly relevant to the EUR/USD exchange rate. To further refine the dataset and extract the relevant segments, we employ ChatGPT-4.0 again to process the text data, creating a final textual dataset consisting of 20,329 texts.
We employ ChatGPT-4.0 to annotate the sentiment polarity scores for the textual training dataset by integrating prompt engineering techniques. To safeguard against potential and unidentified errors, the model is also required to provide explanations for the polarity scores it assigns. In our prompt engineering, we define the polarity score range as [-1,1], where scores approaching 1 indicate a strongly positive sentiment, and vice versa. Scores near zero represent a neutral sentiment.
In terms of next-day exchange rate movement, e.g., up or down, \( M_d \) is used to annotate the text data. \( M_d \) is defined as:
\begin{equation}
M_d = \begin{cases} 
0, & CP_{d+1} < CP_d, \\
1, & CP_{d+1} \geq CP_d,
\end{cases}
\end{equation}
where \( CP_{d-1} \) is the closing price of the exchange rate on trading day \(d+1\) and \( CP_d \) is the closing price on day \(d\). We do not introduce an additional label for \( CP_{d+1} = CP_d \), as it is rare for the closing prices to be the same on two consecutive transaction days.
\subsubsection{RoBERTa-Large}
\begin{figure}[h]
\centering
\includegraphics[width=\linewidth, keepaspectratio]{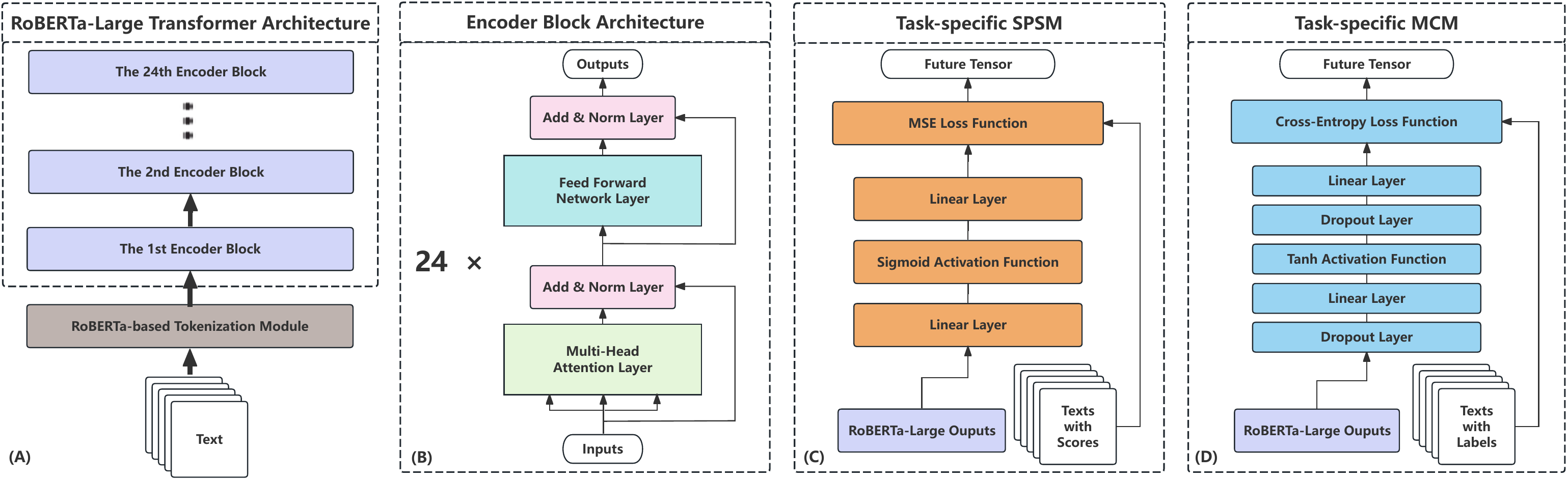} 
\caption{\label{fig:two_module} Components of the RoBERTa-Large Model, including (A) RoBERTa-Large Transformer Architecture, (B) The Encoder Block Architecture, (C) The Sentiment Polarity Scoring Module, and (D) The Movement Classification Module.}
\end{figure}

RoBERTa-Large, developed by Facebook AI, is a SOTA pre-trained LLM incorporating several key enhancements to improve its training process and architecture \cite{DBLP:journals/corr/abs-1907-11692}. Utilizing increased training data, larger batch sizes, and extended training periods, RoBERTa-Large has demonstrated outstanding performance across a variety of benchmark tasks, showcasing its robust capabilities. Its architecture, as illustrated in Figure \ref{fig:two_module}(A), is based on the transformer model and primarily consists of a tokenization module and 24 encoder blocks. Each encoder block, detailed in Figure \ref{fig:two_module}(B), includes a multi-head self-attention mechanism followed by a feed-forward neural network. The self-attention mechanism allows the model to focus on different positions within the input sequence, capturing the relationships and dependencies among tokens.
One of the primary advantages of RoBERTa-Large is its ability to learn robust and transferable language representations. By pre-training on extensive unlabelled text data, RoBERTa-Large develops a deep comprehension of language structure and semantics. As the input flows through each encoding layer, data move from the bottom to the top of the model, with representations becoming increasingly improved and enriched, thereby constructing hierarchical and contextualized embeddings of the text. Therefore, in our SPSM and MCM, we employ two RoBERTa-Large models as the embedding generators. Each text is processed by the RoBERTa-based tokenizer and then fed into the Encoder layer (\(\textit{encoder}(\cdot)\)), generating the model's hidden states (\(T\)):
\begin{equation}
T = \textit{encoder}(\text{[CLS]}, W_1, W_2, \ldots, W_t, \ldots, \text{[PAD]}, \text{[SEP]}).
\end{equation}
Here, \(T \in \mathbb{R}^{1 \times 1024}\), \(W_i\) represents the \(i\)-th word in the text, and \(t\) denotes the length of the text. Positions not occupied by input words are filled with the [\text{PAD}] token to maintain a uniform sequence length of 512. The embeddings (\(e\)) extracted from the \text{CLS} to \text{SEP} positions of \(T\) as the text embeddings:
\begin{equation}
e = T(\text{[CLS] to [SEP]}),
\end{equation}
where \(e\) is a \(1 \times 1024\) vector that passes through all transformer encoder blocks of RoBERTa-Large (\(\textit{RL}(\cdot)\)), resulting in the final output hidden state \(F\):
\begin{equation}
F = \textit{RL}(e).
\end{equation}
Here, \(F \in \mathbb{R}^{1 \times 1024}\), the final hidden state will enter different modules to generate various feature tensors.
\subsubsection{Sentiment Polarity Scoring Module}
In SPSM, our RoBERTa-Large model utilizes weights from the Twitter-RoBERTa-Large-2022-154m model, which is fine-tuned by the CardifNLP team on a large dataset containing 154 million tweets \cite{loureiro2023tweet}. As shown in Figure \ref{fig:two_module}(C), when using the RoBERTa-Large model for sentiment analysis, we expanded the model architecture with a module designed for regression. This module includes a Sigmoid activation function, two linear layers, and a mean squared error (MSE) loss function. The output of the RoBERTa-Large model initially passes through the first linear layer (\(\textit{LL1}_{P}(\cdot)\)), which reduces the high-dimensional text representation from the final hidden state \(F\) of 1024 dimensions to a lower-dimensional space. A Sigmoid activation function (\(\textit{Sig}(\cdot)\)) then compresses this output to a range between 0 and 1, and this output is further transformed by the second linear layer (\(\textit{LL2}_{P}(\cdot)\)) to produce the final feature tensor \(S_{fx}\):
\begin{equation}
S_{fx} = \textit{LL2}_{P}(\textit{Sig}(\textit{LL1}_{P}(F))).
\end{equation}
Here, \(S_{fx} \in \mathbb{R}^{d \times h \times 128}\), \(d\) represents the total number of trading days and \(h\) indicates the maximum number of texts on all trading days.
During training, the predicted sentiment score and the annotated polarity scores are fed into the MSE loss function \cite{ke2019dual,qi2020mean,muthukumar2021classification}.
 The MSE loss function is defined as:
\begin{equation}
L_{MSE} = \frac{1}{n} \sum_{i=1}^{n} (y_i - \hat{y}_i)^2,
\end{equation}
where \(y_i\) is the annotated polarity score for the \(i\)-th text in the dataset, and \(\hat{y}_i\) is the polarity score predicted by the model for the \(i\)-th text. \(n\) is the total number of texts over the training period. By minimizing the MSE loss function, the model learns the mapping relationship between text and sentiment scores. The backpropagation algorithm is used to compute gradients and update the parameters of RoBERTa-Large and SPSM, continually refining the predicted sentiment scores to approach the annotated scores. In addition, we discover that when using the Twitter-RoBERTa-Large-2022-154m model, convergence is faster, and performance on evaluation metrics is superior compared to the base model without fine-tuning, given the same number of training epochs.
\subsubsection{Movement Classification Module}
In MCM, we employ the RoBERTa-Large-Base model to uncover hidden patterns between textual information and the EUR/USD exchange rate movement on the following day \cite{DBLP:journals/corr/abs-1907-11692}. As illustrated in Figure \ref{fig:two_module}(D), we augment the model architecture with a module specifically designed for classification, consisting of five layers. The final hidden state first goes through a dropout layer (\(\textit{DP1}(\cdot)\)) which serves for regularization by randomly masking some neurons during training, thus reducing the risk of overfitting. After this, the output is processed by a linear layer (\(\textit{LL1}_{M}(\cdot)\)), which linearly transforms the hidden state \(F\) from 1024 dimensions to a smaller dimension of 128. This reduction in dimensions decreases the number of parameters and enhances computational efficiency. We then obtain an intermediate representation \(F'\):
\begin{equation}
F' = \textit{LL1}_{M}(\textit{DP1}(F)),
\end{equation}
where \(F'\) refers to a vector of size \(1 \times 128\). A \textit{Tanh} activation function (\(\textit{Tanh}(\cdot)\)) introduces non-linearity, enhancing the model's expressive power and mapping \(F'\) to the range of \([-1,1]\). Another dropout layer (\(\textit{DP2}(\cdot)\)) is added to further regularize the model. The final linear layer (\(\textit{LL2}_{M}(\cdot)\)) maps the intermediate representations to feature tensor \(M_{fx}\), which is defined as:
\begin{equation}
M_{fx} = \textit{LL2}_{M}(\textit{DP2}(\textit{Tanh}(F'))),
\end{equation}
where \(M_{fx} \in \mathbb{R}^{d \times h \times 128}\), \(d\) represents the total number of trading days and \(h\) indicates the maximum number of texts on all transaction days.

In the model training, the Cross-Entropy (CE) loss function measures  the difference between the model's predicted probability distribution and the true labels \cite{hui2020evaluation,mao2023cross,leng2022polyloss}, which can be represented as:
\begin{equation}
L_{CE} = -\frac{1}{n} \sum_{i=1}^{n} \left[ y_i \log(p_i) + (1 - y_i)\log(1 - p_i) \right],
\end{equation}
where \(y_i\) is the true exchange rate movement label for the \(i\)-th text in the dataset, and \(p_i\) is the predicted probability of this movement by the model for the \(i\)-th text. \(n\) is the total number of texts over the training period. During the training process, the model learns the relationship between texts and exchange rate movements by minimizing the cross-entropy loss function, using the backpropagation algorithm to update the parameters of RoBERTa-Large-Base and MCM.
\subsubsection{Experience Rule}
We also collect a financial indicator dataset, which primarily includes data related to EUR/USD exchange rates and financial markets, sourced mainly from financial platforms such as investing.com and finance.yahoo.com, among others. We utilize a comprehensive approach by extensively collecting and analyzing relevant literature to identify the indicators that potentially predict fluctuations in the target exchange rate. The construction of this financial indicator system, displayed in Table \ref{tab:financial indicators}, is based on the evaluation of the relationships between these indicators and the target exchange rate, considering their leading, lagging, and potential non-linear relationships.
\begin{table}[h]
\centering
\begin{tabular}{p{4.5cm}cll}
\hline
\textbf{Classification} & \textbf{No.} & \textbf{Indicator Name} & \textbf{Reference} \\
\hline
Target Series & 1 & EUR/USD Exchange Rate & \\
\hline
US Exchange Rate & 2 & USD/CAD Exchange Rate & \cite{yang2024weighted,barunik2017asymmetric,greenwood2021measuring,kilic2017contagion} \\
(Top 5 Trading Partners) & 3 & USD/MXN Exchange Rate & \\
 & 4 & USD/CNY Exchange Rate & \\
 & 5 & USD/JPY Exchange Rate & \\
 & 6 & USD/KRW Exchange Rate & \\
\hline
EU Exchange Rate & 7 & EUR/CNY Exchange Rate & \cite{yang2024weighted,barunik2017asymmetric,greenwood2021measuring,kilic2017contagion} \\
(Top 5 Trading Partners) & 8 & EUR/GBP Exchange Rate & \\
 & 9 & EUR/CHF Exchange Rate & \\
 & 10 & EUR/RUB Exchange Rate & \\
 & 11 & EUR/TRY Exchange Rate & \\
\hline
Currency Index & 12 & US Dollar Index & \cite{alkan2022currency,ding2021conditional,chantarakasemchit2020forex,gurrib2016optimizing} \\
 & 13 & Euro Index & \cite{alkan2022currency,ding2021conditional,chantarakasemchit2020forex,gurrib2016optimizing} \\
\hline
Currency Futures & 14 & US Dollar Futures-Jun & \cite{tornell2012speculation,ferraro2015can,chen2003commodity} \\
 & 15 & Euro Futures-Jun & \cite{lyons1997simultaneous} \\
\hline
Commodities & 16 & Crude Oil WTI Futures & \cite{cashin2004commodity,jadidzadeh2017does} \\
 & 17 & Natural Gas Futures & \cite{lizardo2010oil,basher2016impact}  \\
 & 18 & Gold Futures & \cite{pukthuanthong2011gold,sjaastad2008price} \\
 & 19 & Copper Futures & \cite{zhang2016exchange,ferraro2015can} \\
 & 20 & Corn Futures & \cite{nazlioglu2012oil,akanni2020returns,nazlioglu2013volatility,rezitis2015relationship}  \\
 & 21 & Soybeans Futures & \cite{nazlioglu2012oil,akanni2020returns,nazlioglu2013volatility,rezitis2015relationship} \\
\hline
Bond Yield & 22 & US 10-Year Bond Yield & \cite{engel2023forecasting,chinn2004monetary,lace2015determining} \\
 & 23 & Eurozone 10-Year Bond Yield & \cite{afonso2018euro,cecioni2018ecb} \\
\hline
Interbank Offered Rate & 24 & SOFR - 1 month & \cite{tonzer2015cross,ivashina2015dollar,dal2012short,duffie2015reforming,du2018deviations} \\
 & 25 & EURIBOR - 1 month & \cite{tonzer2015cross,ivashina2015dollar,dal2012short,eisenschmidt2018measuring} \\
\hline
US Stock Index & 26 & Dow Jones Industrial Average & \cite{pan2007dynamic,tsai2019predict,pandey2018review,nieh2001dynamic,lin2012comovement,phylaktis2005stock,inci2014dynamic,moore2014dynamic,tsai2012relationship} \\
 & 27 & S\&P 500 &  \\
\hline
EU Stock Index & 28 & Euro Stoxx 50 & \cite{pan2007dynamic,tsai2019predict,pandey2018review,nieh2001dynamic,lin2012comovement,phylaktis2005stock,inci2014dynamic,moore2014dynamic,tsai2012relationship} \\
 & 29 & STOXX 600 &  \\
\hline
US Stock Index Futures & 30 & Dow Jones Futures - Jun & \cite{tah2021dynamic,agrawal2010study,andreou2013stock} \\
\hline
EU Stock Index Futures & 31 & EURO STOXX 50 Futures - Jun & \cite{tah2021dynamic,agrawal2010study,andreou2013stock} \\
\hline
Chicago Board Options Exchange & 32 & VIX & \cite{brunnermeier2008carry,cairns2007exchange,pan2019improving} \\
\hline
\end{tabular}
\caption{Comprehensive list of financial indicators used for analysis.}
\label{tab:financial indicators}
\end{table}

All collected raw financial data are fed into an indicator generator, which aligns the data and fills in missing values, producing the final quantitative features. We use linear interpolation to fill in these gaps:
\begin{equation}
v_i = v_a + \frac{(v_b - v_a)(t_i - t_a)}{t_b - t_a},
\end{equation}
here, \( v_i \) represents the interpolated value at the specific time \( t_i \). \( v_a \) and \( v_b \) are the known values at the time points \( t_a \) and \( t_b \), respectively. These known values are used to estimate \( v_i \), assuming a linear change between \( t_a \) and \( t_b \).
\subsection{Causality-Driven Feature Generator }
\begin{figure}[h]
\centering
\includegraphics[width=\linewidth, keepaspectratio]{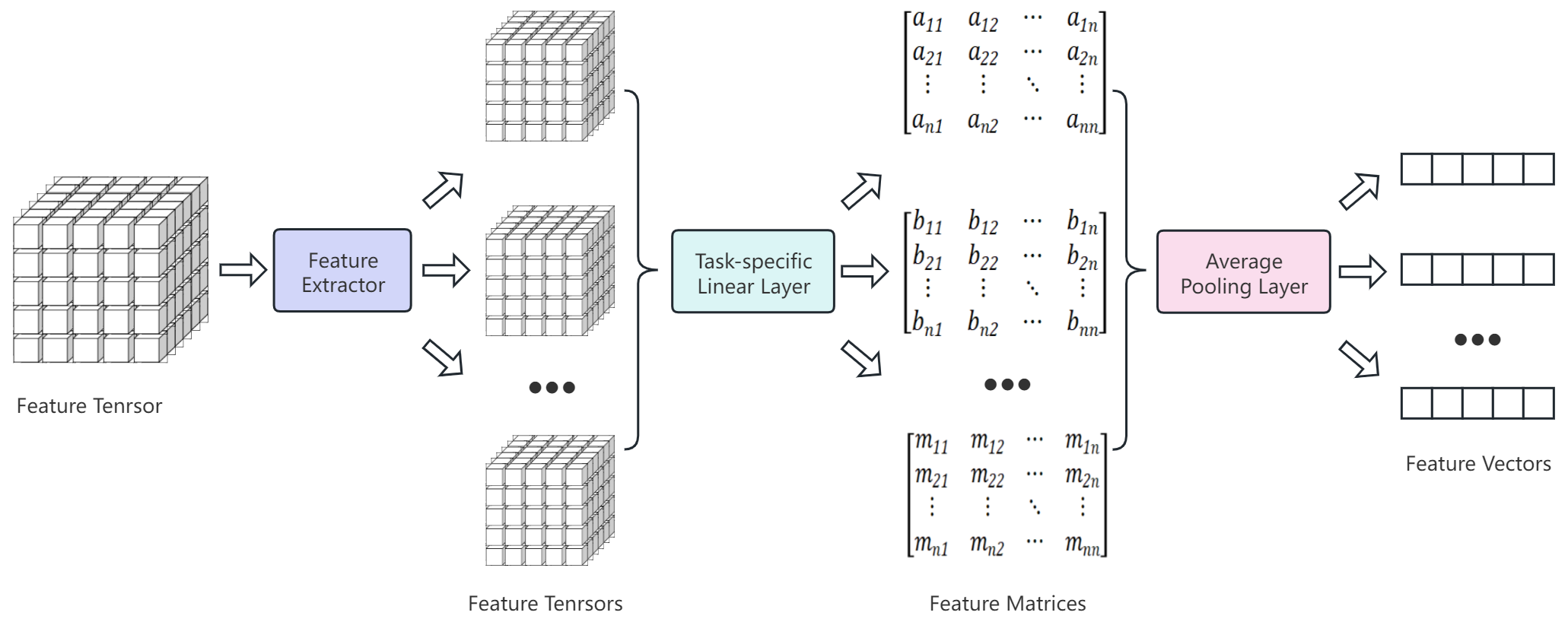} 
\caption{\label{fig:Causality-Driven} The overview of extracting textual features.}
\end{figure}

We employ a Causality-Driven Feature Generator to extract text, exchange rate, and financial market features. Specifically, Figure \ref{fig:Causality-Driven} shows the stage of textual feature extraction, the feature tensor is fed into a feature extractor and produces various types of feature tensors. These produced tensors have the same dimensions as the original feature tensor, which are then processed through a task-specific linear layer, mapping the three-dimensional feature tensors to feature matrices. As for the stage of feature generation, the feature matrices are inputted into an average pooling layer to yield a diverse set of textual features. Subsequently, all textual features combined with other features undergo feature selection to obtain the final set of features inputted into the forecasting model.
\subsubsection{Causality-Driven Feature Generator }
\begin{figure}[h]
\centering
\includegraphics[width=\linewidth, keepaspectratio]{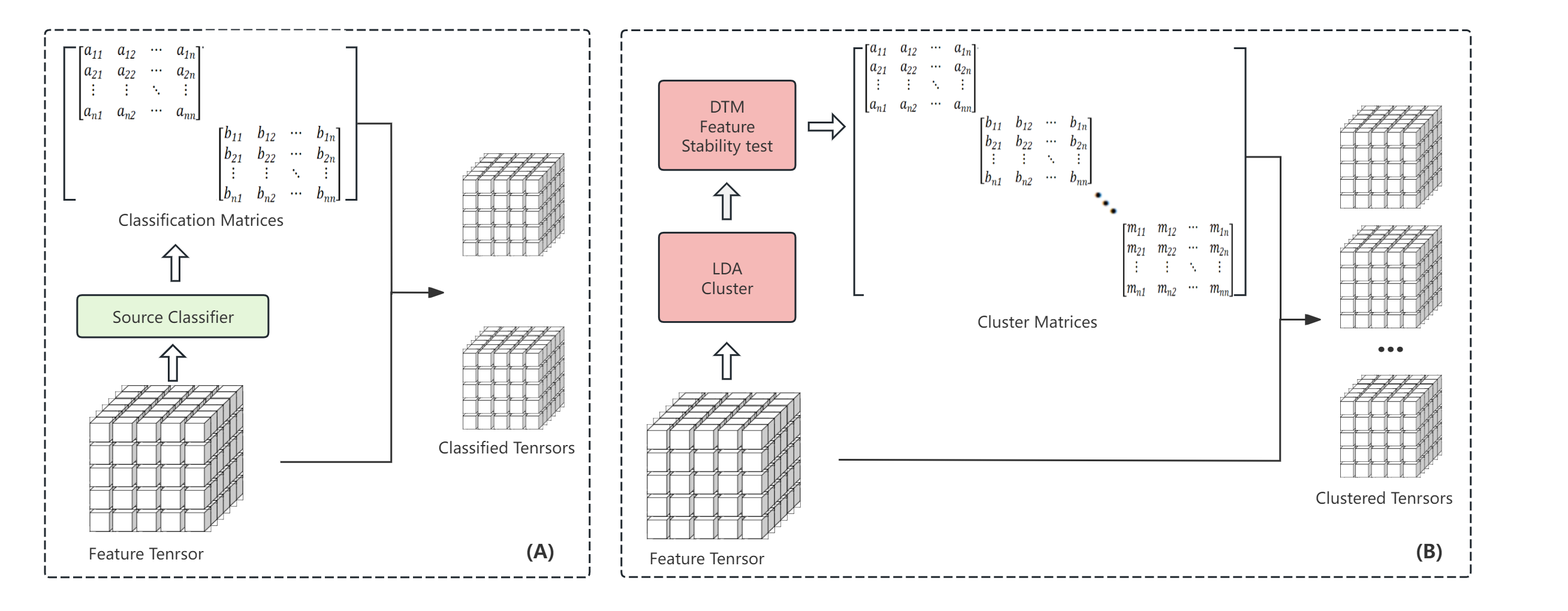} 
\caption{\label{fig:Extractor} The feature extractors include (A)  The source classifier, and (B) The LDA cluster and DTM feature stability test.}
\end{figure}

In terms of feature extraction, we employ two types of feature extractors. The first one is a classifier based on information sources, inspired by the research of Angeletos et al. \cite{angeletos2020business} and Ke et al. \cite{ke2019predicting}. We construct a classifier for news and analysis texts. The news texts encompass a broad spectrum of content related to exchange rates, including central bank monetary policies, economic data releases, geopolitical events, and other significant factors influencing exchange rate trends. In contrast, the analysis texts primarily focus on interpreting, forecasting, and providing recommendations concerning exchange rate movements, typically authored by professional analysts or traders, characterized by their expertise and forward-looking nature. This feature classifier, as depicted in Figure \ref{fig:Extractor}(A), generates news text classification matrix $A$ and analysis text classification matrix $B$. An element of 1 in these matrices indicates that the corresponding vector in the feature tensor is selected by the classifier, and an element of 0 indicates it is not part of that category. The feature tensor is then multiplied by the corresponding matrices to yield the classified feature tensors.

The second extractor is a text-based topic cluster. We utilize an LDA model to cluster the feature tensor, as shown in Figure \ref{fig:Extractor}(B), assigning each text's feature vector to a specific topic\cite{blei2003latent,aziz2022machine}. Furthermore, the optimal number of topics are determined by locating the peak in the topic coherence curve. To cautiously assess the impact of the number of topics on our research, we visualize the outputs of the LDA topic model using the pyLDAvis tool \cite{merter2023evaluation,zhou2023guided}. This tool displays the distance between topics through multidimensional scaling and lists the top significant terms for each topic. To validate the stability of the extracted features, we also conduct dynamic topic modeling (DTM) to examine if the clustered features vary over time \cite{blei2006dynamic,li2019text}. During the feature extraction process, we obtain a series of topics, Topic 1, Topic 2, ..., Topic $n$, and the corresponding text cluster matrices $A$, $B$, ..., $K$. An element of 1 in these matrices indicates that the corresponding vector in the feature tensor is selected by the cluster; an element of 0 indicates it is not part of that category. The feature tensor is then multiplied by the corresponding matrices to yield the clustered feature tensors.
\subsubsection{Feature Generation}
When generating features, we need to handle two types of features: \(S_{fx}\) and \(M_{fx}\). In terms of \(S_{fx}\), after obtaining the sub-tensors from the feature extractor \(\textit{FE}(\cdot)\), we input these sub-tensors into a linear layer \(\textit{LLS}(\cdot)\) specifically designed for processing \(S_{fx}\), mapping them to the corresponding sentiment feature sub-matrix, \(M_s\), each element of which is compressed and normalized to range between [-1,1]:
\begin{equation}
[M_{s1}, M_{s2}, \ldots, M_{sn}] = \textit{LLS}(\textit{FE}(S_{fx})).
\end{equation}
Here, \(M_{si} \in \mathbb{R}^{d \times h}\) is the \(i\)-th feature sub-tensor obtained from \(S_{fx}\) through the feature extractor and subsequently compressed by a linear layer into the corresponding feature matrix, and \(n\) denotes the number of feature sub-tensors of the feature \(S_{fx}\).

Each sentiment feature sub-matrix serves as the input for an average pooling layer, resulting in the corresponding feature vector \(V_s\):
\begin{equation}
V_{si} = \frac{1}{h} \sum_{j=1}^{h} M_{sij}.
\end{equation}
Here, \(M_{si}\) denotes the feature sub-matrix, \(h\) is the number of dimensions in \(M_{si}\). \(M_{sij}\) represents the \(j\)-th column of \(M_{si}\), and \(V_{si} \in \mathbb{R}^{1 \times d}\) is the resultant vector obtained by averaging each dimension of \(M_{si}\).

As for \(M_{fx}\), after obtaining the sub-tensors from the feature extractor \(\textit{FE}(\cdot)\), we input these sub-tensors into a linear layer \(\textit{LLM}(\cdot)\) specifically designed for processing \(M_{fx}\). This process maps each high-dimensional feature vector into a one-dimensional element, which corresponds to two categories in a binary classification task: one category for 'increase and no change', and another for 'decrease'. This mapping forms the corresponding movement feature sub-matrix, \(M_w\):
\begin{equation}
[M_{w1}, M_{w2}, \ldots, M_{wn}] = \textit{LLM}(\textit{FE}(M_{fx})).
\end{equation}
Here, \(M_{wi} \in \mathbb{R}^{d \times h}\) is the \(i\)-th feature sub-tensor obtained from \(M_{fx}\) through the feature extractor and subsequently compressed by a linear layer into the corresponding feature matrix, and \(n\) denotes the number of feature sub-tensors of the feature.

Each exchange movement feature sub-matrix serves as the input for an average pooling layer, resulting in the corresponding feature vector \(V_m\):
\begin{equation}
V_{mi} = \frac{1}{k} \sum_{j=1}^{k} M_{wij}.
\end{equation}
Here, \(M_{wi}\) denotes the feature sub-matrix, \(k\) is the number of dimensions in \(M_{wi}\). \(M_{wij}\) represents the \(j\)-th column of \(M_{wi}\), and \(V_{mi} \in \mathbb{R}^{1 \times d}\) is the vector obtained by averaging each dimension of \(M_{wi}\).

\begin{figure}[h]
\centering
\includegraphics[width=0.45\linewidth, keepaspectratio]{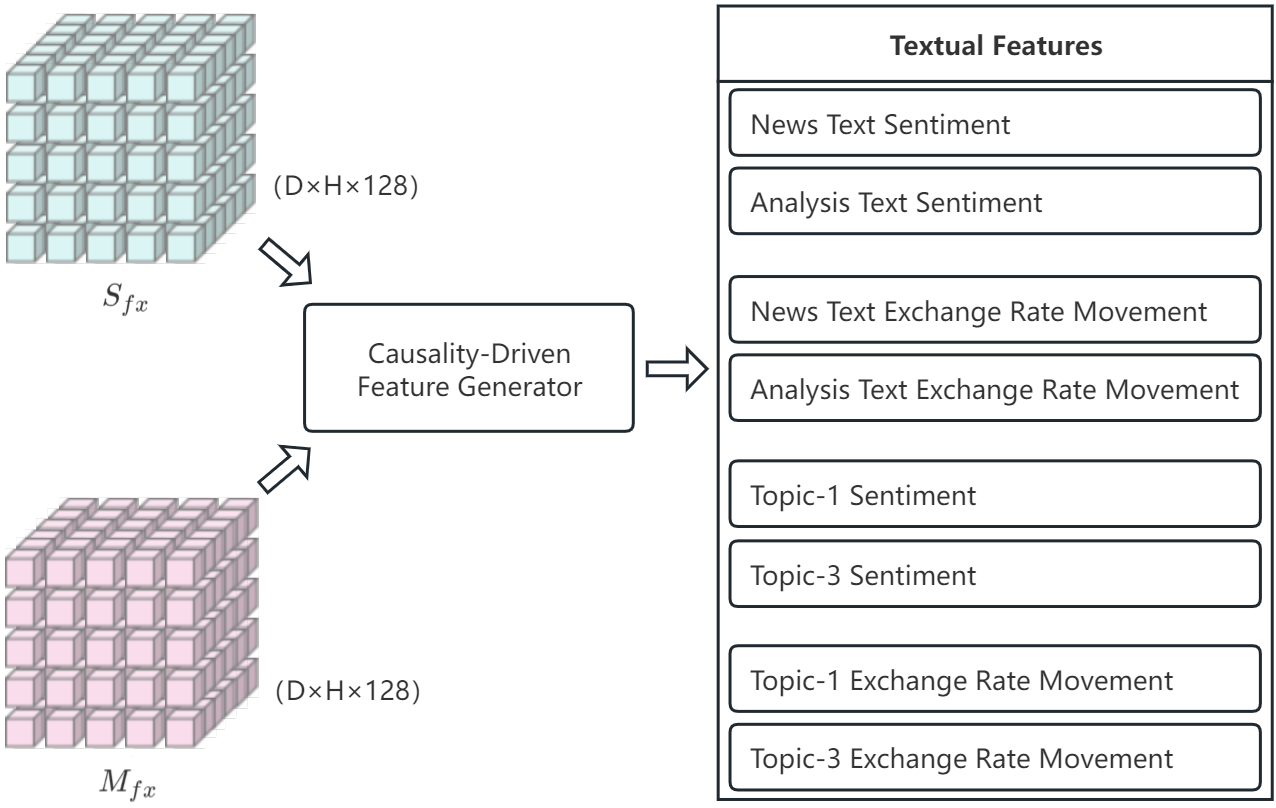} 
\caption{\label{fig:textual-feature}The final textual features generated by the Causality-Driven Feature Generator.}
\end{figure}

Overall, \(S_{fx}\) and \(M_{fx}\) are fed into a Causality-Driven Feature Generator and finally generates 8 textual features as shown in Figure \ref{fig:textual-feature}.
\subsubsection{Feature Selection}
We employ the VAR model to determine the optimal lag orders for all features within the final prediction feature set. The VAR model is an econometric model used to learn the dynamic relationships among multiple time-series variables, expressing each endogenous variable's current value as a linear combination of its own and all other endogenous variables' lagged values \cite{ivanov2005practitioner,stock2001vector}. The VAR(p) model can be represented as follows:
\begin{equation}
Y_t = c + A_1Y_{t-1} + A_2Y_{t-2} + \ldots + A_pY_{t-p} + \epsilon_t.
\end{equation}
Here, \(Y_t\) is an \(n \times 1\) vector of endogenous variables, \(c\) is an \(n \times 1\) vector of constants, \(A_i\) are \(n \times n\) coefficient matrices, and \(\epsilon_t\) is an \(n \times 1\) vector of error terms satisfying the white noise condition. To determine the optimal lag order \(p\), we use the \textit{Akaike Information Criterion} (\textit{AIC}):
\begin{equation}
\text{AIC}(p) = \ln \left( \det(\hat{\Sigma}_p) \right) + \frac{2pn^2}{T}.
\end{equation}
Where, \(\hat{\Sigma}_p\) is the estimated covariance matrix of the residuals for \(p\) lags, indicating the overall variance that the model fails to explain with smaller values being preferable. The natural logarithm of the determinant of this matrix, \(\ln(\det(\hat{\Sigma}_p))\), quantifies the total unexplained variance by the model, serving as a measure of model fit. The penalty term, \(\frac{2pn^2}{T}\), increases with the number of lags (\(p\)) and the number of endogenous variables (\(n\)), normalized by the sample size (\(T\)), to penalize model complexity and prevent overfitting, promoting a balance between fitting accuracy and model simplicity \cite{musa2024application,mondal2023multivariate,schorfheide2018identifying}.

In order to select the optimal lag order, \textit{VAR} models are estimated at different lag orders for each feature, ranging from 0 to 10 to encompass potential fluctuations. Corresponding \textit{AIC} values are then calculated to determine the optimal lag for each feature. The lag order associated with the lowest \textit{AIC} value is considered optimal, as a lower \textit{AIC} value indicates a better balance between model fit and complexity\cite{bai2022crude,li2019text}. Table \ref{tab:optimal lag} displays the optimal lag periods for all features.
\begin{table}[ht]
\centering
\resizebox{1\linewidth}{!} {
\begin{tabular}{@{}lllllllllllll@{}}
\toprule
\textbf{Feature Name} & \textbf{Lag} & \textbf{Feature Name} & \textbf{Lag} & \textbf{Feature Name} & \textbf{Lag} \\ \midrule
EUR/USD ER & 2 & Euro Futures-Jun & 3 & STOXX 600 & 1 \\
USD/CAD ER & 2 & Crude Oil WTI Futures & 1 & EURO STOXX 50 Futures - Jun & 1 \\
USD/MXN ER & 1 & Natural Gas Futures & 1 & Dow Jones Futures - Jun & 1 \\
USD/CNY ER & 1 & Gold Futures & 1 & VIX & 1 \\
USD/JPY ER & 1 & Copper Futures & 1 & News Text Sentiment & 4 \\
USD/KRW ER & 1 & Corn Futures & 1 & Analysis Text Sentiment & 1 \\
EUR/CNY ER & 1 & Soybeans Futures & 1 & News Text ER Movement & 4 \\
EUR/GBP ER & 1 & US 10-Year Bond Yield & 3 & Analysis Text ER Movement & 1 \\
EUR/CHF ER & 1 & Eurozone 10-Year Bond Yield & 2 & Topic-1 Sentiment & 4 \\
EUR/RUB ER & 1 & SOFR - 1 Month & 1 & Topic-3 Sentiment & 1 \\
EUR/TRY ER & 1 & EURIBOR - 1 Month & 1 & Topic-1 ER Movement & 5 \\
US Dollar Index & 3 & Dow Jones Industrial Average & 1 & Topic-3 ER Movement & 3 \\
Euro Index & 7 & S\&P 500 & 1 & & \\
US Dollar Futures-Jun & 3 & Euro Stoxx 50 & 1 & & \\ \bottomrule
\end{tabular}
}
\vspace{1ex}

\footnotesize{\textbf{Note:} ER = Exchange Rate}
\caption{The optimal lag periods for all features}
\label{tab:optimal lag}
\end{table}
After adjusting all features to their optimal lag orders, we employ the Recursive Feature Elimination (RFE) method with random forest regression to rank the importance of all indicators. RFE is a backward elimination algorithm that recursively removes the least important features until the desired number of features is determined \cite{darst2018using,zhou2016cost,gregorutti2017correlation}. Random forest regressor, an ensemble of decision trees each trained on a bootstrap sample of the training data, determines the importance of each feature by averaging the decrease in impurity this feature causes across all trees in the forest. This decrease in impurity, also known as Gini importance or mean decrease impurity, enhances prediction accuracy and minimizes overfitting. The formula for calculating the importance of a feature \( f \), denoted as \( I(f) \), is defined as:
\begin{equation}
I(f) = \sum_{t=1}^{T} \sum_{n=1}^{N_t} \Delta i(n, f).
\end{equation}
Where \( I(f) \) is the importance of feature \( f \), \( T \) is the number of trees in the forest, \( N_t \) is the number of nodes in tree \( t \), and \( \Delta i(n, f) \) is the impurity decrease caused by feature \( f \) at node \( n \). The RFE algorithm initiates by training a random forest regressor on the initial set of features, calculating feature importances, removing the least important features, and repeating these steps until the desired number of features is retained. The RFE process can be expressed as:
\begin{equation}
F_i = F_{i-1} \setminus \{f_j\},
\end{equation}
where \( F_i \) is the feature set at iteration \( i \), \( F_{i-1} \) is the feature set from the previous iteration, and \( f_j \) is the least important feature removed at iteration \( i \). This recursive removal of less important features and evaluation of model performance ultimately helps identify an optimal feature subset containing the most important features.

We use \( FS(\cdot) \) to denote feature selection and all features obtained through the Causality-Driven Feature Generator are concatenated to form the input feature set \( I \):
\begin{equation}
I = FS([V_{s1}, \ldots, V_{sn}] \oplus [V_{m1}, \ldots, V_{mn}] \oplus [V_{e1}, \ldots, V_{en}] \oplus [V_{f1}, \ldots, V_{fn}]).
\end{equation}
Where, \( I \in \mathbb{R}^{112 \times 470} \) represents the feature set, including 111 features over 470 trading days, comprising 80 financial features and 31 textual features. The symbol \( \oplus \) denotes the concatenation operation. Finally, before inputting the selected features into the predictive model, we perform min-max normalization to scale all features to the range of [0, 1]. The normalization is applied as follows:
\begin{equation}
I'_i = \frac{I_i - \min(I_i)}{\max(I_i) - \min(I_i)},
\end{equation}
where \( I'_i \) represents the \( i \)-th row of the matrix \( I \), corresponding to the \( i \)-th feature across all dates. \( I'_i \) is the normalized \( i \)-th row of the matrix, min (\( I_i \)) and max (\( I_i \)) are the minimum and maximum values found in the \( i \)-th row of \( I \). After normalization, rows \( I'_1 \) to \( I'_{112} \) are concatenated to form the final feature matrix \( I' \in \mathbb{R}^{112 \times 470} \), which is then used as the input for the forecasting model.

\subsection{Optuna-Bi-LSTM}
\subsubsection{Bi-LSTM}
\begin{figure}[!htb]
\centering
\includegraphics[width=\linewidth, keepaspectratio]{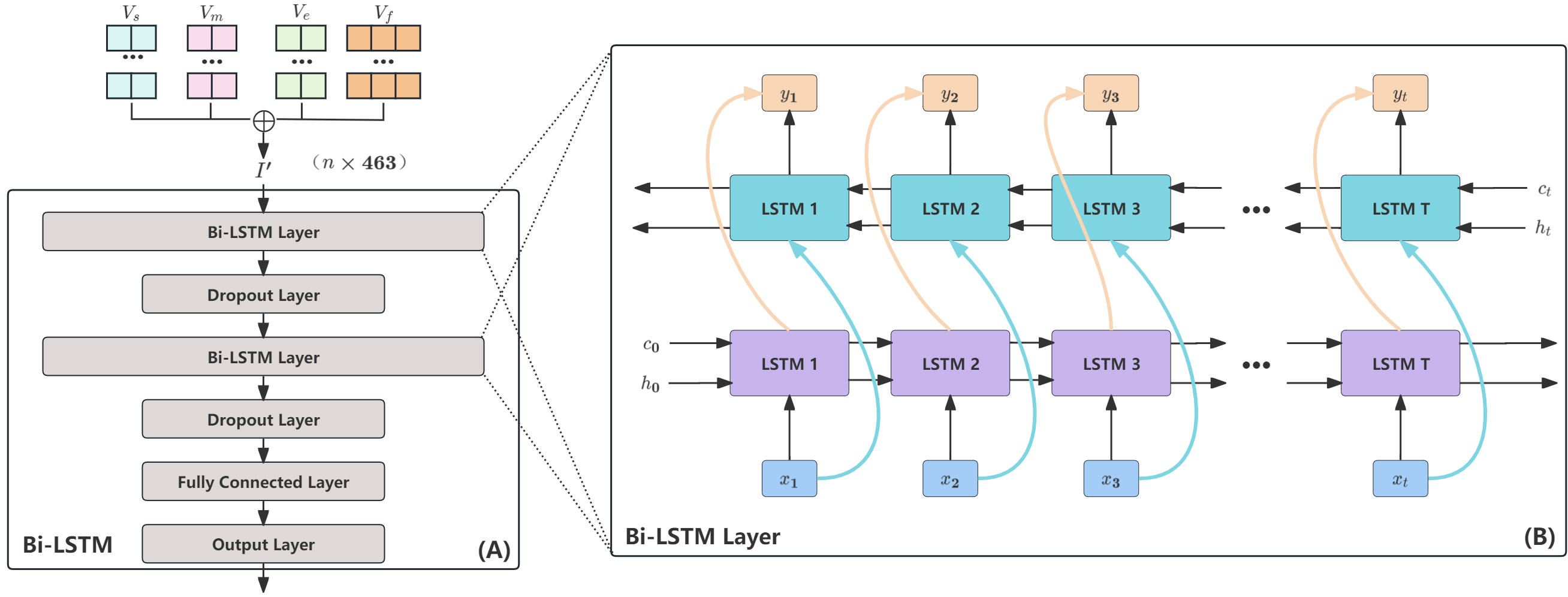} 
\caption{The Bi-LSTM model includes: (A) Bi-LSTM architecture and (B) the workflow of Bi-LSTM Layer.}
\label{fig:Bi-LSTM}
\end{figure}

This study employs a Bi-LSTM model to analyze financial features and forecast EUR/USD foreign exchange rates. The Bi-LSTM is an efficient sequential learning model that enhances performance by integrating past and future feature information, demonstrating strong capabilities in time series forecasting \cite{yang2022adaptability,ma2023multi,siami2019performance}. Figure \ref{fig:Bi-LSTM}(A) illustrates the structure of our prediction model, where \( I' \) is inputted into two Bi-LSTM layers, \textit{Bi-LSTM1}(\(\cdot\)) and \textit{Bi-LSTM2}(\(\cdot\)), to identify temporal patterns for predicting EUR/USD exchange rate movements. The key to the model lies in the Bi-LSTM layers, which capture both forward and backward dependencies in the input time sequence. Figure \ref{fig:Bi-LSTM}(B) displays the structure of each Bi-LSTM layer, composed of a forward LSTM and a backward LSTM, processing the sequential information in both directions respectively. After each Bi-LSTM layer, a dropout layer, \textit{DP1}(\(\cdot\)) and \textit{DP2}(\(\cdot\)), is added to utilize regularization to reduce the risk of overfitting. The final hidden state from \( I' \) at the last time step, \( h_{t+1} \), can be expressed as:
\begin{equation}
h_{t+1} = \textit{DP2} \left( \textit{Bi-LSTM2} \left( \textit{DP1} \left( \textit{Bi-LSTM1} (I') \right) \right) \right),
\end{equation}
where \( h_{t+1} \in \mathbb{C}^{1 \times w} \), \( c1 \) is the number of feature combinations in the current model, and \( w \) is the sliding window length. \( h_{t+1} \) carries all the features necessary for predicting the EUR/USD exchange rate movement on day \( t+1 \). Finally, \( h_{t+1} \) passes through a fully connected layer (\textit{FC}(\(\cdot\))) and an output layer (\textit{Output}(\(\cdot\))), generating the predicted EUR/USD exchange rate \( \hat{y}_{t+1} \) for day \( t+1 \):
\begin{equation}
\hat{y}_{t+1} = \textit{Output}(\textit{FC}(h_{t+1})).
\end{equation}
We utilize the MSE loss function to measure the discrepancy between predicted values and actual values, updating the model parameters accordingly:
\begin{equation}
L_{\text{MSE}} = \frac{1}{n} \sum_{i=1}^{n} (y_{t+i} - \hat{y}_{t+i})^2,
\end{equation}
where \( y_{t+1} \) represents the actual price range of the EUR/USD exchange rate on day \( t+1 \). By minimizing the loss function, the model learns the relationship between the features within the rolling window and the subsequent trading day's EUR/USD exchange rate.
\subsubsection{Optuna optimization framework}

Optuna is an automatic hyperparameter optimization framework that efficiently searches for the optimal set of hyperparameters \cite{akiba2019optuna,zhou2024ship}. Its core principle is to generate trials, each testing a different combination of hyperparameters, while using the results of previous trials to guide the sampling of parameters in subsequent trials. This sampling, combined with the ability to prune underperforming trials, allows Optuna to quickly converge on the best hyperparameter configuration.

\begin{figure}[!htb]
\centering
\includegraphics[width=0.45\linewidth, keepaspectratio]{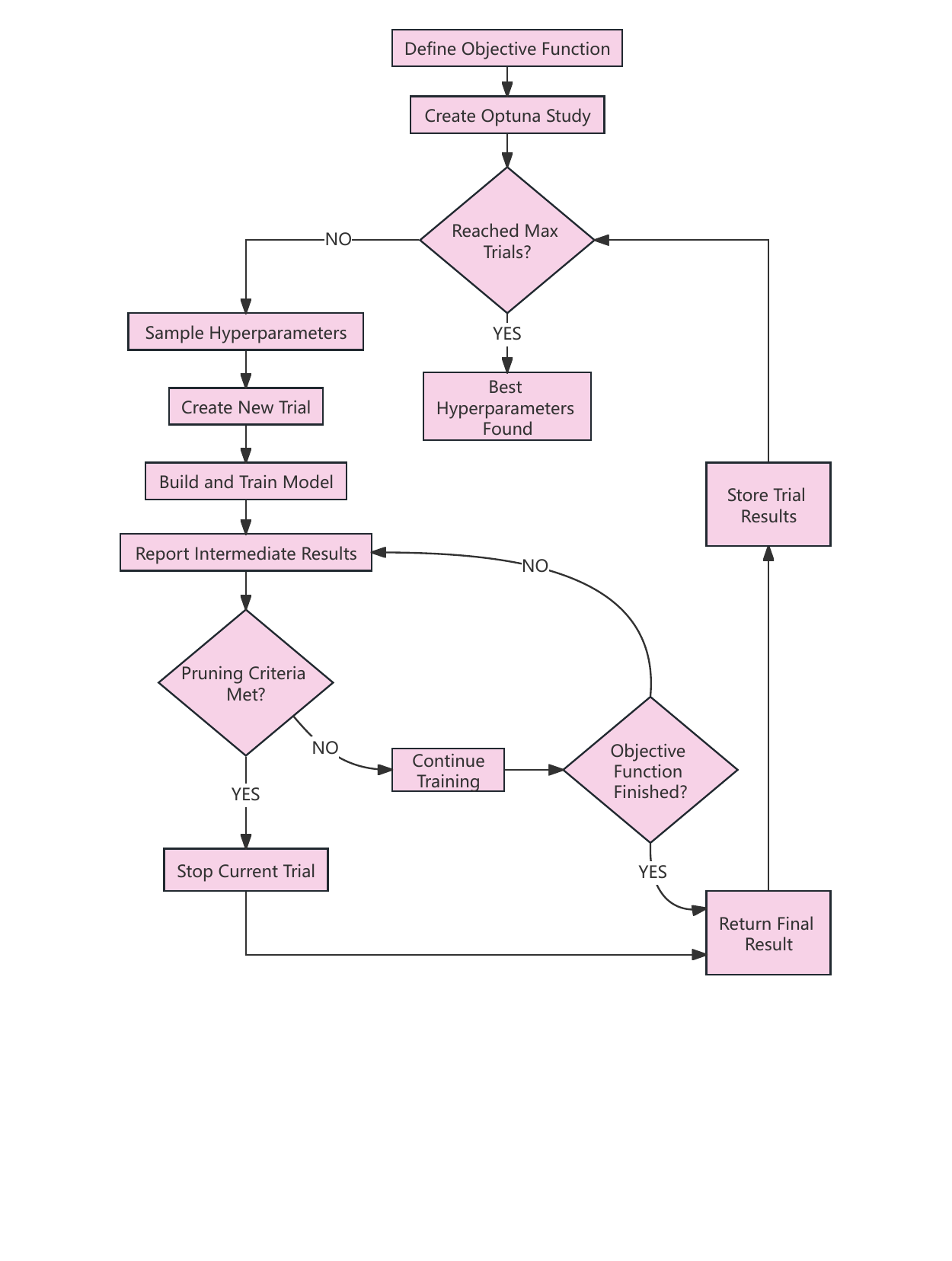} 
\caption{\label{fig:optuna}The workflow of the Optuna hyperparameter optimization process.}
\end{figure}

The Figure \ref{fig:optuna} illustrates the process of hyperparameter optimization using Optuna. The process begins by defining an objective function for constructing, training, and evaluating the model. Subsequently, an Optuna Study object is created to manage the optimization process, and the optimization loop then begins, generating new hyperparameters based on the results of previous trials. Each trial is represented by a Trial object and passed to the objective function. Inside the objective function, the suggest method of the Trial object is utilized to obtain hyperparameter values. The model is built and trained using these hyperparameters, and intermediate results are reported by the Trial's report method. If the pruning criteria are met, the trial is terminated; otherwise, training continues until the objective function is completed. The final evaluation result is returned, and the trial results are stored. This process is repeated until the maximum number of trials is reached. Finally, the best hyperparameter combination is obtained from the Study object.

The parameter space in this study consists of six dimensions, shown in Table \ref{tab:parameter-space}. Among these, hidden layer size, fully connected layer size, batch size, and window size are discrete variables, but window size also includes continuous ranges within specific numbers. Dropout rate and learning rate are continuous variables, with the learning rate being sampled on a logarithmic scale. Additionally, the size of the fully connected layer is constrained by the size of the hidden layer and must be less than or equal to the hidden layer size.

\begin{table}[!htbp]
\centering
\begin{tabular}{@{}lc@{}}
\toprule
\textbf{Parameter}                  & \textbf{Range}                                \\ \midrule
Hidden layer size                   & [8, 16, 32, 64, 128]                          \\
Fully connected layer size          & [8, 16, 32, 64] \\
Dropout rate                        & [0.0, 0.5]                                    \\
Learning rate                       & [1e-5, 1e-2]                     \\
Batch size                          & [8, 16, 32, 48, 64, 80, 96, 112, 128]                          \\
Window size (continuous)            & [1, 24]                                       \\
Window size (discrete)              & [30, 40, 50, 60]                              \\ \bottomrule
\end{tabular}
\caption{The hyperparameter space.}
\label{tab:parameter-space}
\end{table}

\section{Experiment}
\subsection{Dataset}
\begin{figure}[!htbp]
\centering
\includegraphics[width=0.6\linewidth, keepaspectratio]{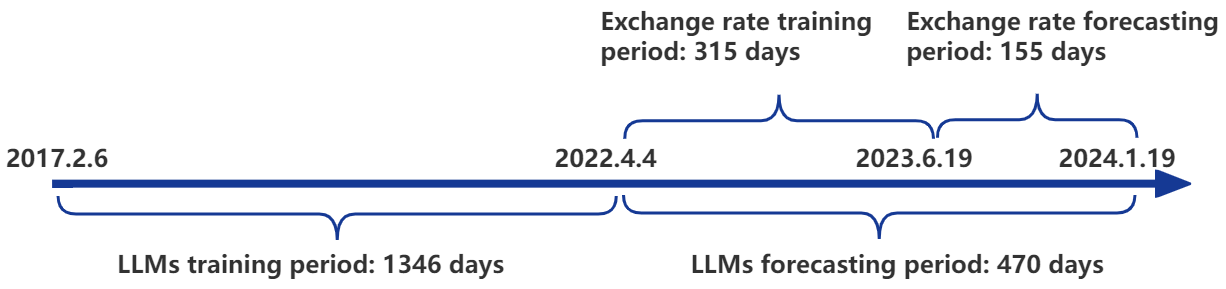} 
\caption{\label{fig:time line}Training and testing periods of the two LLMs and the exchange rate forecasting models.}
\end{figure}

During the training phase of two LLMs, the datasets span from February 6, 2017, to April 4, 2022. Subsequently, the two fine-tuned LLMs are utilized to score text sentiment polarity and classify exchange rate movements in the predictive dataset ranging from April 4, 2022, to January 19, 2024. As for the exchange rate prediction models, their training set consist of the first 315 transaction days from April 4, 2022, to January 19, 2024, and their prediction set comprises the following 155 transaction days, as shown in Figure \ref{fig:time line}. To prevent information leakage, we implement strict data processing methods, ensuring that the predictive data remains isolated from the training process.

\subsection{Evaluation metrics}
We evaluate the models’ forecasting performances over the test period using two different criteria: the mean absolute error (MAE) and root mean squared error (RMSE):
\begin{equation}
\text{MAE} = \frac{1}{n} \sum_{i=1}^n |\hat{y}_i - y_i|,
\end{equation}
\begin{equation}
\text{RMSE} = \sqrt{\frac{1}{n} \sum_{i=1}^n (\hat{y}_i - y_i)^2}.
\end{equation}
Here, \(\hat{y}_i\) represents the model's forecasted exchange rate for day \(i\), \(y_i\) represents the actual exchange rate for day \(i\), and \(n\) is the total number of days in the testset \cite{lei2021volatility}.

To quantify the additional explanatory power of the textual features, we compute a metric named the percentage improvement (PI) . This metric represents the percentage enhancement in predictive performance achieved by combining the two feature sets compared to using the quantitative financial features alone \cite{li2019text}. The two PIs are defined as follows:
\begin{equation}
PI_{\text{MAE}} = \left(\frac{\text{MAE of Quantitative Features} - \text{MAE of Combined Features}}{\text{MAE of Quantitative Features}}\right) \times 100\%
\label{eq:mae_improvement}
\end{equation}
\begin{equation}
PI_{\text{RMSE}} = \left(\frac{\text{RMSE of Quantitative Features} - \text{RMSE of Combined Features}}{\text{RMSE of Quantitative Features}}\right) \times 100\%
\label{eq:rmse_improvement}
\end{equation}

\section{Experiment Results}
\subsection{Main Results}
\subsubsection{Time series forecasting}

We compare the performance of the Optuna-Bi-LSTM model with other benchmark models. The prediction results of different models for the EUR/USD exchange rate are shown in Table~\ref{tab:model-comparison}. It can be observed that our proposed method consistently outperforms other models in terms of both MAE and RMSE metrics, achieving an improvement of at least 10.69\% in MAE and 9.56\% in RMSE compared with the best other model.

\begin{figure}[!htbp]
\centering
\includegraphics[width=\linewidth, keepaspectratio]{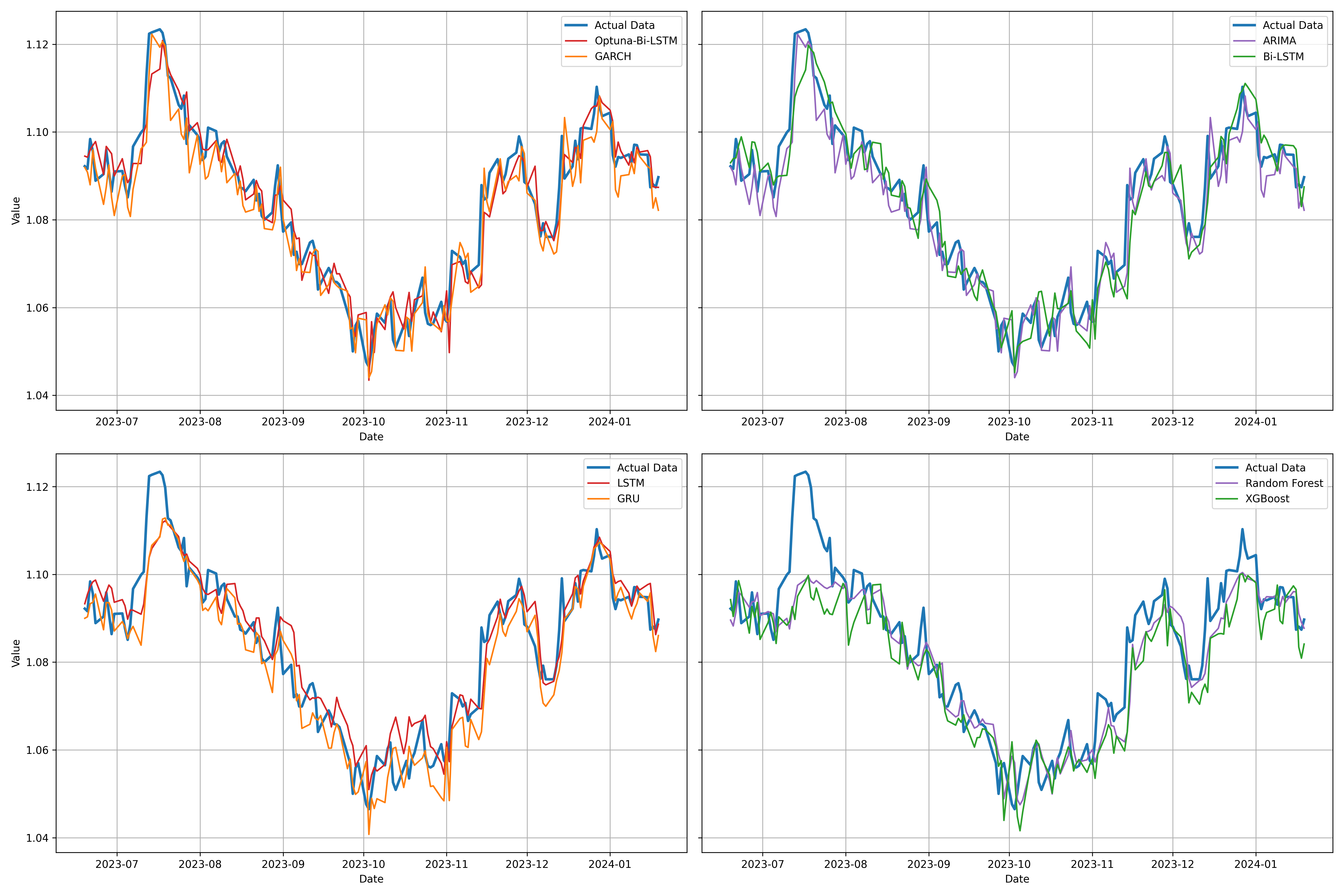} 
\caption{\label{fig:time forecasting line}Training and testing periods of the two LLMs and the exchange rate forecasting models.}
\end{figure}

Moreover, as illustrated in Figure~\ref{fig:time forecasting line}, the prediction curve obtained by Optuna-Bi-LSTM aligns more closely with the raw data curve and exhibits a higher degree of trend similarity. This demonstrates the superior predictive performance of our proposed model.

\begin{table}[!htbp]
\centering
\begin{tabular}{@{}llllll@{}}
\toprule
Series        & Category                & Model                   & MAE       & RMSE      & Rank \\ \midrule
Multivariate  & Proposed method         & Optuna-Bi-LSTM         & 0.003746  & 0.004982  & 1    \\
              & Deep learning methods   & Bi-LSTM                 & 0.004511  & 0.005814  & 4    \\
              &                         & LSTM                    & 0.004768  & 0.006212  & 5    \\
              &                         & GRU                     & 0.004958  & 0.006457  & 6    \\
              & Machine learning methods& Random Forest           & 0.005471  & 0.007672  & 7    \\
              &                         & XGBoost                 & 0.006809  & 0.009012  & 8    \\
Univariate    & Statistical methods     & GARCH                   & 0.004282  & 0.005695  & 2    \\
              &                         & ARIMA                   & 0.004456  & 0.005718  & 3    \\ \bottomrule
\end{tabular}
\caption{Comparison of Forecasting Models based on MAE and RMSE}
\label{tab:model-comparison}
\end{table}

The results indicate that by combining structured and unstructured data for exchange rate prediction and feeding the features generated by the Causality-Driven Feature Generator into the model, our proposed method achieves significant improvements over the benchmarks.

\subsubsection{DM Test}
To evaluate and compare the predictive performance of the eight models on the time series, we conducted the Diebold-Mariano (DM) test. The purpose of the DM test is to determine which models are statistically significantly superior to others in terms of prediction accuracy \cite{diebold2002comparing,molodtsova2009out,rossi2013exchange}. We first calculated the forecast errors for each model and constructed error difference series from them for pairwise comparisons between models. Using this approach, we performed a total of 28 DM tests, covering all possible combinations of model pairs. The test results are summarized in Table~\ref{tab:model_ranking}, where the LSTM model demonstrates the best performance, while the GRU model exhibits the poorest predictive ability.

\begin{table}[!htb]
\centering
\begin{tabular}{cl}
\toprule
\textbf{Rank} & \textbf{Model} \\
\midrule
1 & LSTM \\
2 & Optuna-Bi-LSTM \\
3 & Bi-LSTM \\
4 & Random Forest \\
5 & XGBoost \\
6 & GARCH \\
7 & ARIMA \\
8 & GRU \\
\bottomrule
\end{tabular}
\caption{Ranking of forecasting models Based on DM Test Results.}
\label{tab:model_ranking}
\end{table}

\subsubsection{Window Size Analysis}
\begin{figure}[!htbp]
\centering
\includegraphics[width=\linewidth, keepaspectratio]{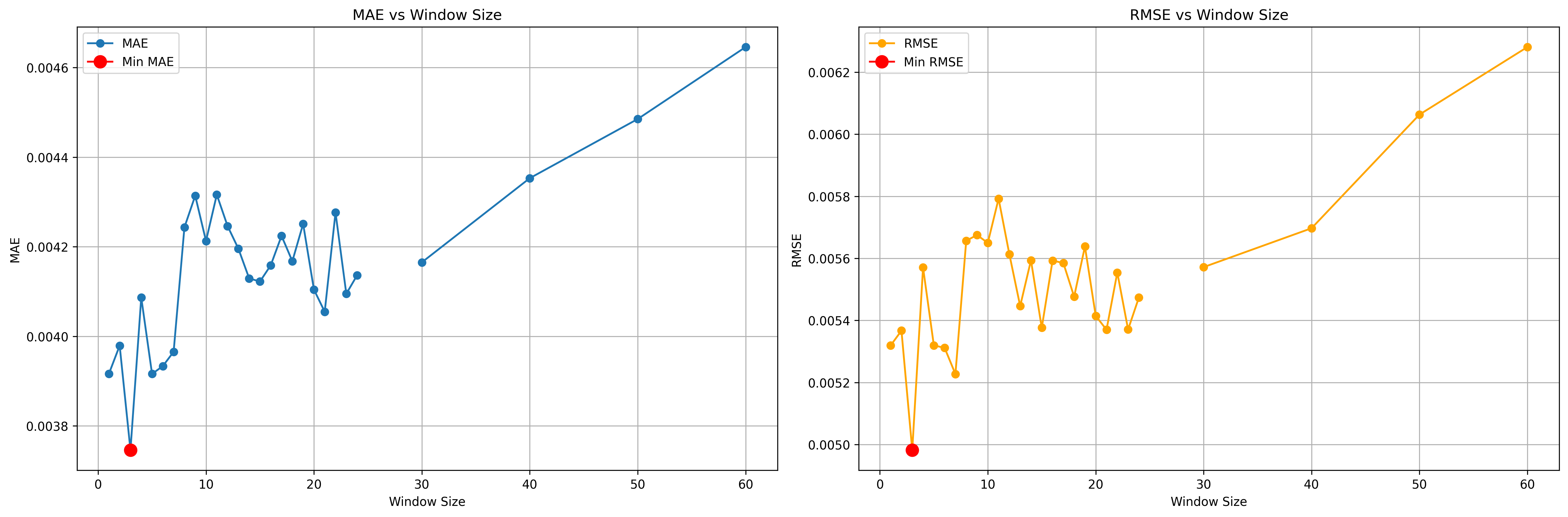} 
\caption{\label{fig:window size}Performance metrics varying with window size.}
\end{figure}

In this study, we analyze the impact of different window sizes on the prediction accuracy of the models. The tested window sizes range from 1 to 24, with additional extended sizes of 30, 40, 50, and 60, to assess their influence on model performance. This analysis helps determine the optimal window size that maximizes the accuracy of time series predictions. The results of this analysis are presented in Figure~\ref{fig:window size}. By observing the changes in MAE and RMSE across different models, each window size is evaluated. When the window size is 3, the models exhibit the best performance. As the window size increases, the model performance generally deteriorates.

\subsection{Ablation Experiment}
\subsubsection{Textual Feature Breakdown}
We investigate the relative predictive power of textual and other features by comparing the performance of using (1) only textual features (31 features), (2) only exchange rate and financial market features (the top 12 important features selected by the RFE method), and all selected features (43 features), which is the combination of (1) and (2). The results show that in the scenario where only textual features are inputted, the model's performance is far inferior to the scenarios using only financial features and all selected features. This indicates that textual features alone may not provide sufficient predictive information for exchange rate forecasting. Moreover, when using Optuna-Bi-LSTM, the combination of textual and financial features achieves the best performance, outperforming the case of using only financial features.
\begin{table}[!htbp]
\centering
\begin{tabular}{lcccc}
\hline
 & \multicolumn{1}{c}{(1)} & \multicolumn{1}{c}{(2)} & \multicolumn{1}{c}{Combination: (1) + (2)} & \multicolumn{1}{c}{PI from (2) to (1) + (2)} \\ \hline
MAE & & & & \\ \hline
Optuna-Bi-LSTM       &  0.025406                              & 0.004270                                      & 0.003746                                     & 12.27\%                                                          \\
Bi-LSTM                 & 0.020712                                  & 0.004736                                      & 0.004511                                     & 4.75\%                                                           \\
Random Forest   & 0.031247                                 & 0.005747                                      & 0.005471                                      & 4.80\%                                                         \\ \hline
RMSE & & & & \\ \hline
Optuna-Bi-LSTM       & 0.029987                                & 0.005502                                     & 0.004982                                      & 9.45\%                                                           \\
Bi-LSTM                & 0.024674                                  & 0.005816                                      & 0.005814                                      & 0.03\%                                                           \\
Random Forest   & 0.032789                                 & 0.007813                                      & 0.007672                                     & 1.8\%                                                         \\ \hline
\end{tabular}
\caption{Forecasting performances with different features.}
\label{tab:Textual Feature Breakdown}
\end{table}

The last column of Table~\ref{tab:Textual Feature Breakdown} presents the percentage improvement results defined by Formulas~\ref{eq:mae_improvement} and \ref{eq:rmse_improvement}. Whether in terms of MAE or RMSE, the improvement of combined features over financial features alone in the Optuna-Bi-LSTM model significantly outperforms that of the Bi-LSTM model. The overall prediction results suggest that textual and financial features are complementary to each other and are suitable for exchange rate prediction. When unstructured data and structured data are combined in a predictive model, significant accuracy improvements can be achieved.

\subsubsection{Breakdown Study}

We conduct several ablation experiments to analyze the effectiveness of each textual feature generated by the IUS framework. In each experiment, we include the exchange rate and financial market features (the top 12 important features selected by the RFE method) as a fixed set of predictors. The textual feature sets we evaluate include: (A) features obtained through SSM and Source Classification, (B) features obtained through SSM and LDA Cluster, (C) features obtained through MCM and Source Classification, and (D) features obtained through MCM and LDA Cluster. We use (A + B + C + D) to represent the complete set of predictive features, including both textual and financial features, and (0) to represent no textual features, only the fixed set of financial features.

\begin{table}[!htbp]
    \centering
    \begin{tabular}{lcccccc}
        \toprule
        \textbf{Textual features} & \textbf{MAE} & \textbf{Rank (MAE)} & \textbf{RMSE} & \textbf{Rank (RMSE)} & \textbf{Weighted Rank} \\
        \midrule
        (0) & 0.004270 & 16 & 0.005502 & 16 & 16 \\
        (A) & 0.003495 & 1 & 0.004716 & 1 & 1 \\
        (B) & 0.003975 & 12 & 0.005164 & 10 & 10 \\
        (C) & 0.003945 & 10 & 0.005115 & 6 & 8 \\
        (D) & 0.004095 & 15 & 0.005342 & 15 & 15 \\
        (A)+(B) & 0.003634 & 2 & 0.004879 & 2 & 2 \\
        (A)+(C) & 0.003843 & 6 & 0.005015 & 4 & 5 \\
        (A)+(D) & 0.003773 & 4 & 0.005062 & 5 & 4 \\
        (B)+(C) & 0.003889 & 8 & 0.005144 & 9 & 9 \\
        (B)+(D) & 0.004021 & 13 & 0.005270 & 13 & 14 \\
        (C)+(D) & 0.003861 & 7 & 0.005121 & 7 & 7 \\
        (A)+(B)+(C) & 0.004024 & 14 & 0.005207 & 11 & 13 \\
        (A)+(B)+(D) & 0.003809 & 5 & 0.005131 & 8 & 6 \\
        (A)+(C)+(D) & 0.003917 & 9 & 0.005287 & 14 & 11 \\
        (B)+(C)+(D) & 0.003969 & 11 & 0.005268 & 12 & 12 \\
        (A)+(B)+(C)+(D) & 0.003746 & 3 & 0.004982 & 3 & 3 \\
        \bottomrule
    \end{tabular}
    \caption{Performance metrics for different textual feature combinations}
    \label{tab:performance_metrics}
\end{table}
In Table \ref{tab:performance_metrics}, we calculate the weighted rank as follows: \(\text{Weighted Score} = 0.5 \times \text{MAE Rank} + 0.5 \times \text{RMSE Rank}\). When the ranks are tied, we prioritize the MAE rank for ascending order. As shown in the table, using only (A) and (A)+(B) results in lower MAE and RMSE compared to our proposed method.

\begin{figure}[!htb]
    \centering
    \includegraphics[width=\textwidth]{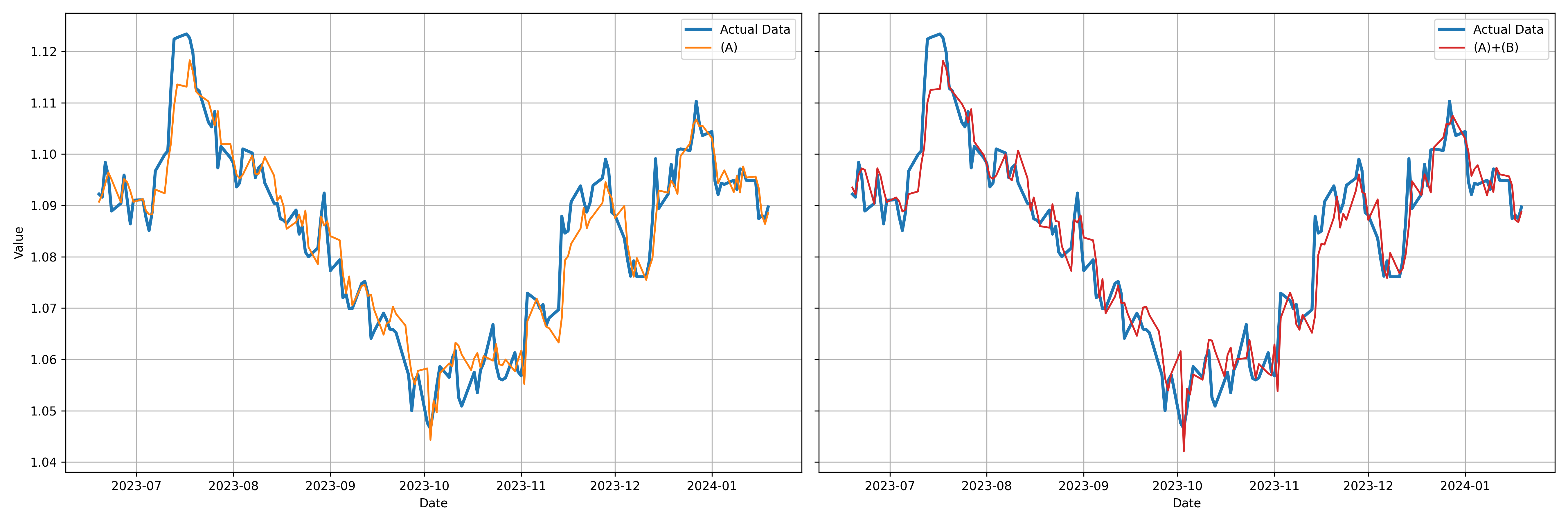}
    \caption{Prediction curves using textual features (A) and (A)+(B) for predicting exchange rates.}
    \label{fig:textual_features}
\end{figure}

Figure \ref{fig:textual_features} presents the prediction curves when using only (A) and (A)+(B) to predict the exchange rates. We observe that the two prediction curves are closer to the original data curve compared to our proposed method. Therefore, when using more features for prediction, the additional textual features (C)+(D) may introduce slight noise.

\subsubsection{Recursive Factor Importance Feature Selection}

We select different numbers of exchange rate and financial related features based on their importance, along with all textual features, to input into the predictive model. This approach allows us to observe the impact of the number of features on model predictive performance. 
\begin{figure}[!htb]
\centering
\includegraphics[width=\linewidth, keepaspectratio]{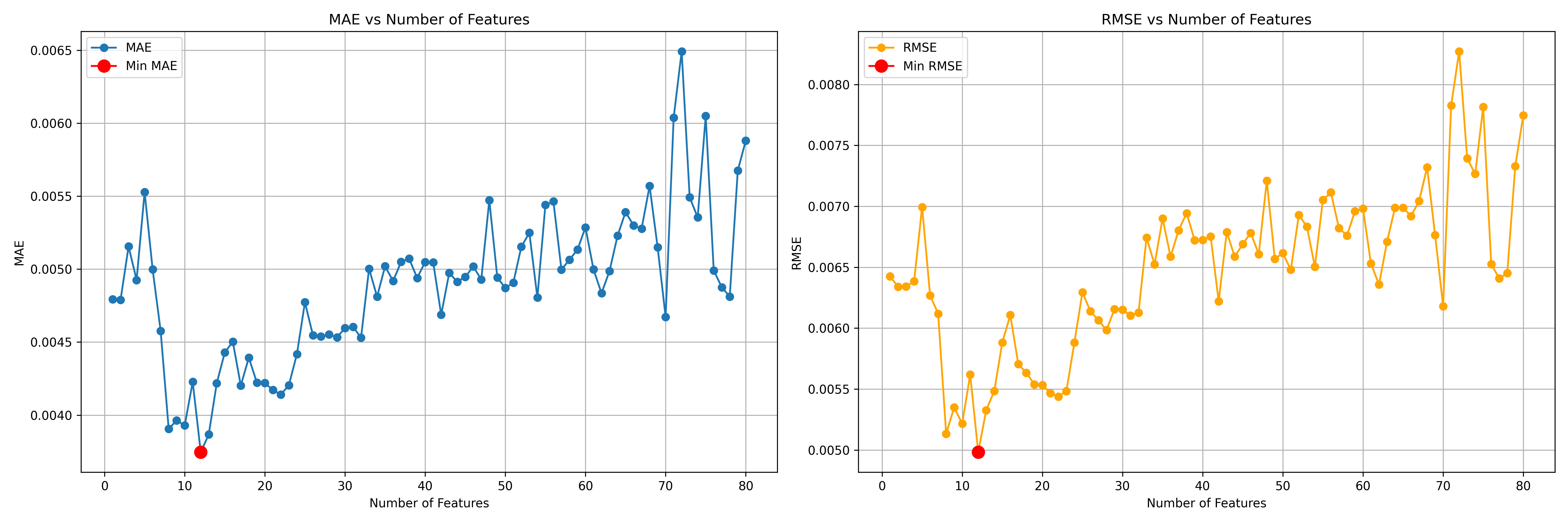} 
\caption{\label{fig:RFE}The workflow of the Optuna hyperparameter optimization process.}
\end{figure}

The Figure \ref{fig:RFE} illustrates the curve of MAE and RMSE as the number of features varies. As the financial features increase from 1 to 12, both curves reach their minimum values. Subsequently, both curves rise steeply. This indicates that as the number of features increases, noise is introduced into the predictive model, thereby gradually reducing its predictive capability. Overall, the top 12 important features selected by the RFE method, combined with 31 textual features, constitute a robust feature set for predicting exchange rates.

\section{Conclusions and Future Work}
To accurately forecast the EUR/USD exchange rate, we introduced the IUS framework and an Optuna-Bi-LSTM model. This framework integrates multi-source information, including news and analytical texts, other relevant exchange rates, and financial market indicators. We employ two large language models for sentiment polarity scoring and exchange rate movement classification, which are then combined with other quantitative indicators input into a Causality-Driven Feature Generator. All generated features are fed into the predictive model for exchange rate forecasting.

Experimental results demonstrate that compared to the strongest benchmarks, our method achieved the highest MAE and RMSE, improving by at least 10.69\% and 9.56\%, respectively. In terms of data fusion, by combining unstructured and structured data, the model is able to enhance prediction accuracy beyond what is possible with structured data alone. Furthermore, using the top 12 important features selected by the RFE method, combined with 31 textual features proves to be more effective compared to analyzing all textual features, as it more directly corresponds to the actual exchange rate response to market conditions. In summary, the proposed method achieves better performance in exchange rate forecasting and provides a comprehensive approach by integrating multi-source data for enhanced prediction accuracy.

\bibliographystyle{unsrtnat}
\bibliography{references}

\begin{thebibliography}{122}
\providecommand{\natexlab}[1]{#1}
\providecommand{\url}[1]{\texttt{#1}}
\expandafter\ifx\csname urlstyle\endcsname\relax
  \providecommand{\doi}[1]{doi: #1}\else
  \providecommand{\doi}{doi: \begingroup \urlstyle{rm}\Url}\fi

\bibitem[Rossi(2013)]{rossi2013exchange}
Barbara Rossi.
\newblock Exchange rate predictability.
\newblock \emph{Journal of economic literature}, 51\penalty0 (4):\penalty0
  1063--1119, 2013.

\bibitem[Cheung et~al.(2019)Cheung, Chinn, Pascual, and
  Zhang]{cheung2019exchange}
Yin-Wong Cheung, Menzie~D Chinn, Antonio~Garcia Pascual, and Yi~Zhang.
\newblock Exchange rate prediction redux: New models, new data, new currencies.
\newblock \emph{Journal of International Money and Finance}, 95:\penalty0
  332--362, 2019.

\bibitem[Sezer et~al.(2020)Sezer, Gudelek, and Ozbayoglu]{sezer2020financial}
Omer~Berat Sezer, Mehmet~Ugur Gudelek, and Ahmet~Murat Ozbayoglu.
\newblock Financial time series forecasting with deep learning: A systematic
  literature review: 2005--2019.
\newblock \emph{Applied soft computing}, 90:\penalty0 106181, 2020.

\bibitem[Hu et~al.(2021)Hu, Zhao, and Khushi]{hu2021survey}
Zexin Hu, Yiqi Zhao, and Matloob Khushi.
\newblock A survey of forex and stock price prediction using deep learning.
\newblock \emph{Applied System Innovation}, 4\penalty0 (1):\penalty0 9, 2021.

\bibitem[Semiromi et~al.(2020)Semiromi, Lessmann, and Peters]{semiromi2020news}
Hamed~Naderi Semiromi, Stefan Lessmann, and Wiebke Peters.
\newblock News will tell: Forecasting foreign exchange rates based on news
  story events in the economy calendar.
\newblock \emph{The North American Journal of Economics and Finance},
  52:\penalty0 101181, 2020.

\bibitem[Singh et~al.(2024)Singh, Rani, Chugh, and Singh]{singh2024utilising}
Harmanjeet Singh, Ashima Rani, Aarti Chugh, and Supreet Singh.
\newblock Utilising sentiment analysis to predict market movements in currency
  exchange rates.
\newblock In \emph{2024 3rd International Conference for Innovation in
  Technology (INOCON)}, pages 1--7. IEEE, 2024.

\bibitem[Tadphale et~al.(2023)Tadphale, Saraswat, Sonawane, and
  Deshmukh]{tadphale2023impact}
Anushka Tadphale, Haripriya Saraswat, Omkar Sonawane, and PR~Deshmukh.
\newblock Impact of news sentiment on foreign exchange rate prediction.
\newblock In \emph{2023 3rd International Conference on Intelligent
  Technologies (CONIT)}, pages 1--8. IEEE, 2023.

\bibitem[Haider et~al.(2023)Haider, Nazir, Jim{\'e}nez, and
  Jibran~Qamar]{haider2023commodity}
Saba Haider, Mian~Sajid Nazir, Alfredo Jim{\'e}nez, and Muhammad~Ali
  Jibran~Qamar.
\newblock Commodity prices and exchange rates: evidence from
  commodity-dependent developed and emerging economies.
\newblock \emph{International Journal of Emerging Markets}, 18\penalty0
  (1):\penalty0 241--271, 2023.

\bibitem[Sarkar and Ali(2022)]{sarkar2022eur}
MS~Aziz Sarkar and UAM~Ehsan Ali.
\newblock Eur/usd exchange rate prediction using machine learning.
\newblock \emph{Int. J. Math. Sci. Comput}, 8\penalty0 (1):\penalty0 44--48,
  2022.

\bibitem[Ruan et~al.(2024)Ruan, Zhang, and Lv]{ruan2024forecasting}
Qingsong Ruan, Jiarui Zhang, and Dayong Lv.
\newblock Forecasting exchange rate volatility: is economic policy uncertainty
  better?
\newblock \emph{Applied Economics}, 56\penalty0 (13):\penalty0 1526--1544,
  2024.

\bibitem[Windsor and Cao(2022)]{windsor2022improving}
Edmure Windsor and Wei Cao.
\newblock Improving exchange rate forecasting via a new deep multimodal fusion
  model.
\newblock \emph{Applied Intelligence}, 52\penalty0 (14):\penalty0 16701--16717,
  2022.

\bibitem[Salisu et~al.(2021)Salisu, Cu{\~n}ado, Isah, and Gupta]{salisu2021oil}
Afees~A Salisu, Juncal Cu{\~n}ado, Kazeem Isah, and Rangan Gupta.
\newblock Oil price and exchange rate behaviour of the brics.
\newblock \emph{Emerging Markets Finance and Trade}, 57\penalty0 (7):\penalty0
  2042--2051, 2021.

\bibitem[Neghab et~al.(2024)Neghab, Cevik, Wahab, and
  Basar]{neghab2024explaining}
Davood~Pirayesh Neghab, Mucahit Cevik, MIM Wahab, and Ayse Basar.
\newblock Explaining exchange rate forecasts with macroeconomic fundamentals
  using interpretive machine learning.
\newblock \emph{Computational Economics}, pages 1--43, 2024.

\bibitem[Ito and Takeda(2022)]{ito2022sentiment}
Takumi Ito and Fumiko Takeda.
\newblock Do sentiment indices always improve the prediction accuracy of
  exchange rates?
\newblock \emph{Journal of Forecasting}, 41\penalty0 (4):\penalty0 840--852,
  2022.

\bibitem[Ben~Omrane et~al.(2020)Ben~Omrane, Welch, and Zhou]{ben2020dynamic}
Walid Ben~Omrane, Robert Welch, and Xinyao Zhou.
\newblock The dynamic effect of macroeconomic news on the euro/us dollar
  exchange rate.
\newblock \emph{Journal of Forecasting}, 39\penalty0 (1):\penalty0 84--103,
  2020.

\bibitem[Li et~al.(2019)Li, Shang, and Wang]{li2019text}
Xuerong Li, Wei Shang, and Shouyang Wang.
\newblock Text-based crude oil price forecasting: A deep learning approach.
\newblock \emph{International Journal of Forecasting}, 35\penalty0
  (4):\penalty0 1548--1560, 2019.

\bibitem[Bai et~al.(2022)Bai, Li, Yu, and Jia]{bai2022crude}
Yun Bai, Xixi Li, Hao Yu, and Suling Jia.
\newblock Crude oil price forecasting incorporating news text.
\newblock \emph{International Journal of Forecasting}, 38\penalty0
  (1):\penalty0 367--383, 2022.

\bibitem[Swathi et~al.(2022)Swathi, Kasiviswanath, and Rao]{swathi2022optimal}
T~Swathi, N~Kasiviswanath, and A~Ananda Rao.
\newblock An optimal deep learning-based lstm for stock price prediction using
  twitter sentiment analysis.
\newblock \emph{Applied Intelligence}, 52\penalty0 (12):\penalty0 13675--13688,
  2022.

\bibitem[Kalamara et~al.(2022)Kalamara, Turrell, Redl, Kapetanios, and
  Kapadia]{kalamara2022making}
Eleni Kalamara, Arthur Turrell, Chris Redl, George Kapetanios, and Sujit
  Kapadia.
\newblock Making text count: economic forecasting using newspaper text.
\newblock \emph{Journal of Applied Econometrics}, 37\penalty0 (5):\penalty0
  896--919, 2022.

\bibitem[Naeem et~al.(2021)Naeem, Mashwani, Ali, Uddin, Mahmoud, Jamal, and
  Chesneau]{naeem2021machine}
Samreen Naeem, Wali~Khan Mashwani, Aqib Ali, M~Irfan Uddin, Marwan Mahmoud,
  Farrukh Jamal, and Christophe Chesneau.
\newblock Machine learning-based usd/pkr exchange rate forecasting using
  sentiment analysis of twitter data.
\newblock \emph{Computers, Materials \& Continua}, 67\penalty0 (3):\penalty0
  3451--3461, 2021.

\bibitem[Xueling et~al.(2023)Xueling, Xiong, and Yucong]{xueling2023exchange}
Lv~Xueling, Xiong Xiong, and Shen Yucong.
\newblock Exchange rate market trend prediction based on sentiment analysis.
\newblock \emph{Computers and Electrical Engineering}, 111:\penalty0 108901,
  2023.

\bibitem[K{\"u}{\c{c}}{\"u}klerli and Ulusoy(2024)]{kuccuklerli2024sentiment}
Kaz{\i}m~Berk K{\"u}{\c{c}}{\"u}klerli and Veysel Ulusoy.
\newblock Sentiment-driven exchange rate forecasting: Integrating twitter
  analysis with economic indicators.
\newblock \emph{Journal of Applied Finance \& Banking}, 14\penalty0
  (3):\penalty0 75--96, 2024.

\bibitem[Li et~al.(2023)Li, Yin, and Zhang]{li2023analysis}
Jiatong Li, Jiawen Yin, and Rui Zhang.
\newblock Analysis and forecast of usd/eur exchange rate based on arima and
  garch models.
\newblock In \emph{International Conference on Economic Management and Green
  Development}, pages 566--575. Springer, 2023.

\bibitem[Zhang(2024)]{zhang2024rmb}
Ruimin Zhang.
\newblock Rmb/usd exchange rate forecasting and analysis by ses and arima
  model.
\newblock \emph{Highlights in Business, Economics and Management}, 24:\penalty0
  112--121, 2024.

\bibitem[Colombo and Pelagatti(2020)]{colombo2020statistical}
Emilio Colombo and Matteo Pelagatti.
\newblock Statistical learning and exchange rate forecasting.
\newblock \emph{International Journal of Forecasting}, 36\penalty0
  (4):\penalty0 1260--1289, 2020.

\bibitem[Pfahler(2021)]{pfahler2021exchange}
Jonathan~Felix Pfahler.
\newblock Exchange rate forecasting with advanced machine learning methods.
\newblock \emph{Journal of Risk and Financial Management}, 15\penalty0
  (1):\penalty0 2, 2021.

\bibitem[Khoa and Huynh(2022)]{khoa2022predicting}
Bui~Thanh Khoa and Tran~Trong Huynh.
\newblock Predicting exchange rate under uirp framework with support vector
  regression.
\newblock \emph{assessment}, 12:\penalty0 13, 2022.

\bibitem[Sun et~al.(2020)Sun, Wang, and Wei]{sun2020new}
Shaolong Sun, Shouyang Wang, and Yunjie Wei.
\newblock A new ensemble deep learning approach for exchange rates forecasting
  and trading.
\newblock \emph{Advanced Engineering Informatics}, 46:\penalty0 101160, 2020.

\bibitem[Liu et~al.(2024)Liu, Huang, Li, and Wei]{liu2024new}
Siyuan Liu, Qiqian Huang, Mingchen Li, and Yunjie Wei.
\newblock A new lasso-bilstm-based ensemble learning approach for exchange rate
  forecasting.
\newblock \emph{Engineering Applications of Artificial Intelligence},
  127:\penalty0 107305, 2024.

\bibitem[Islam and Hossain(2021)]{islam2021foreign}
Md~Saiful Islam and Emam Hossain.
\newblock Foreign exchange currency rate prediction using a gru-lstm hybrid
  network.
\newblock \emph{Soft Computing Letters}, 3:\penalty0 100009, 2021.

\bibitem[Dautel et~al.(2020)Dautel, H{\"a}rdle, Lessmann, and
  Seow]{dautel2020forex}
Alexander~Jakob Dautel, Wolfgang~Karl H{\"a}rdle, Stefan Lessmann, and
  Hsin-Vonn Seow.
\newblock Forex exchange rate forecasting using deep recurrent neural networks.
\newblock \emph{Digital Finance}, 2:\penalty0 69--96, 2020.

\bibitem[Wan et~al.(2019)Wan, Mei, Wang, Liu, and Yang]{wan2019multivariate}
Renzhuo Wan, Shuping Mei, Jun Wang, Min Liu, and Fan Yang.
\newblock Multivariate temporal convolutional network: A deep neural networks
  approach for multivariate time series forecasting.
\newblock \emph{Electronics}, 8\penalty0 (8):\penalty0 876, 2019.

\bibitem[Zeng et~al.(2023)Zeng, Chen, Zhang, and Xu]{zeng2023transformers}
Ailing Zeng, Muxi Chen, Lei Zhang, and Qiang Xu.
\newblock Are transformers effective for time series forecasting?
\newblock In \emph{Proceedings of the AAAI conference on artificial
  intelligence}, volume~37, pages 11121--11128, 2023.

\bibitem[Xu et~al.(2022)Xu, Hu, Wu, Jian, Li, Chen, Zhang, Zhang, and
  Wang]{xu2022research}
Yuanhao Xu, Caihong Hu, Qiang Wu, Shengqi Jian, Zhichao Li, Youqian Chen,
  Guodong Zhang, Zhaoxi Zhang, and Shuli Wang.
\newblock Research on particle swarm optimization in lstm neural networks for
  rainfall-runoff simulation.
\newblock \emph{Journal of hydrology}, 608:\penalty0 127553, 2022.

\bibitem[Hamdia et~al.(2021)Hamdia, Zhuang, and Rabczuk]{hamdia2021efficient}
Khader~M Hamdia, Xiaoying Zhuang, and Timon Rabczuk.
\newblock An efficient optimization approach for designing machine learning
  models based on genetic algorithm.
\newblock \emph{Neural Computing and Applications}, 33\penalty0 (6):\penalty0
  1923--1933, 2021.

\bibitem[Victoria and Maragatham(2021)]{victoria2021automatic}
A~Helen Victoria and Ganesh Maragatham.
\newblock Automatic tuning of hyperparameters using bayesian optimization.
\newblock \emph{Evolving Systems}, 12\penalty0 (1):\penalty0 217--223, 2021.

\bibitem[Dong et~al.(2019)Dong, Shen, Wang, Shao, Ling, and
  Porikli]{dong2019dynamical}
Xingping Dong, Jianbing Shen, Wenguan Wang, Ling Shao, Haibin Ling, and Fatih
  Porikli.
\newblock Dynamical hyperparameter optimization via deep reinforcement learning
  in tracking.
\newblock \emph{IEEE transactions on pattern analysis and machine
  intelligence}, 43\penalty0 (5):\penalty0 1515--1529, 2019.

\bibitem[Amboage et~al.(2024)Amboage, Wulff, Girone, and
  Pena]{amboage2024model}
Juan Pablo~Garc{\'\i}a Amboage, Eric Wulff, Maria Girone, and Tom{\'a}s~F Pena.
\newblock Model performance prediction for hyperparameter optimization of deep
  learning models using high performance computing and quantum annealing.
\newblock In \emph{EPJ Web of Conferences}, volume 295, page 12005. EDP
  Sciences, 2024.

\bibitem[Brodzicki et~al.(2021)Brodzicki, Piekarski, and
  Jaworek-Korjakowska]{brodzicki2021whale}
Andrzej Brodzicki, Micha{\l} Piekarski, and Joanna Jaworek-Korjakowska.
\newblock The whale optimization algorithm approach for deep neural networks.
\newblock \emph{Sensors}, 21\penalty0 (23):\penalty0 8003, 2021.

\bibitem[Das and Srihari(2024)]{das2024compos}
Souvik Das and Rohini~K Srihari.
\newblock Compos mentis at semeval2024 task6: A multi-faceted role-based large
  language model ensemble to detect hallucination.
\newblock In \emph{Proceedings of the 18th International Workshop on Semantic
  Evaluation (SemEval-2024)}, pages 1449--1454, 2024.

\bibitem[Gallipoli et~al.(2024)Gallipoli, Papicchio, Vaiani, Cagliero, Miola,
  and Borghi]{gallipoli2024keyword}
Giuseppe Gallipoli, Simone Papicchio, Lorenzo Vaiani, Luca Cagliero, Arianna
  Miola, and Daniele Borghi.
\newblock Keyword-based annotation of visually-rich document content for trend
  and risk analysis using large language models.
\newblock In \emph{Proceedings of the Joint Workshop of the 7th Financial
  Technology and Natural Language Processing, the 5th Knowledge Discovery from
  Unstructured Data in Financial Services, and the 4th Workshop on Economics
  and Natural Language Processing@ LREC-COLING 2024}, pages 130--136, 2024.

\bibitem[Li et~al.(2024)Li, Xu, Tang, and Wen]{li2024knowledge}
Hang Li, Tianlong Xu, Jiliang Tang, and Qingsong Wen.
\newblock Knowledge tagging system on math questions via llms with flexible
  demonstration retriever.
\newblock \emph{arXiv preprint arXiv:2406.13885}, 2024.

\bibitem[Liu et~al.(2019)Liu, Ott, Goyal, Du, Joshi, Chen, Levy, Lewis,
  Zettlemoyer, and Stoyanov]{DBLP:journals/corr/abs-1907-11692}
Yinhan Liu, Myle Ott, Naman Goyal, Jingfei Du, Mandar Joshi, Danqi Chen, Omer
  Levy, Mike Lewis, Luke Zettlemoyer, and Veselin Stoyanov.
\newblock Roberta: {A} robustly optimized {BERT} pretraining approach.
\newblock \emph{CoRR}, abs/1907.11692, 2019.
\newblock URL \url{http://arxiv.org/abs/1907.11692}.

\bibitem[Loureiro et~al.(2023)Loureiro, Rezaee, Riahi, Barbieri, Neves, Anke,
  and Camacho-Collados]{loureiro2023tweet}
Daniel Loureiro, Kiamehr Rezaee, Talayeh Riahi, Francesco Barbieri, Leonardo
  Neves, Luis~Espinosa Anke, and Jose Camacho-Collados.
\newblock Tweet insights: A visualization platform to extract temporal insights
  from twitter.
\newblock \emph{arXiv preprint arXiv:2308.02142}, 2023.

\bibitem[Ke et~al.(2019{\natexlab{a}})Ke, Wang, Yan, Ren, and Lau]{ke2019dual}
Zhanghan Ke, Daoye Wang, Qiong Yan, Jimmy Ren, and Rynson~WH Lau.
\newblock Dual student: Breaking the limits of the teacher in semi-supervised
  learning.
\newblock In \emph{Proceedings of the IEEE/CVF international conference on
  computer vision}, pages 6728--6736, 2019{\natexlab{a}}.

\bibitem[Qi et~al.(2020)Qi, Du, Siniscalchi, Ma, and Lee]{qi2020mean}
Jun Qi, Jun Du, Sabato~Marco Siniscalchi, Xiaoli Ma, and Chin-Hui Lee.
\newblock On mean absolute error for deep neural network based vector-to-vector
  regression.
\newblock \emph{IEEE Signal Processing Letters}, 27:\penalty0 1485--1489, 2020.

\bibitem[Muthukumar et~al.(2021)Muthukumar, Narang, Subramanian, Belkin, Hsu,
  and Sahai]{muthukumar2021classification}
Vidya Muthukumar, Adhyyan Narang, Vignesh Subramanian, Mikhail Belkin, Daniel
  Hsu, and Anant Sahai.
\newblock Classification vs regression in overparameterized regimes: Does the
  loss function matter?
\newblock \emph{Journal of Machine Learning Research}, 22\penalty0
  (222):\penalty0 1--69, 2021.

\bibitem[Hui and Belkin(2020)]{hui2020evaluation}
Like Hui and Mikhail Belkin.
\newblock Evaluation of neural architectures trained with square loss vs
  cross-entropy in classification tasks.
\newblock \emph{arXiv preprint arXiv:2006.07322}, 2020.

\bibitem[Mao et~al.(2023)Mao, Mohri, and Zhong]{mao2023cross}
Anqi Mao, Mehryar Mohri, and Yutao Zhong.
\newblock Cross-entropy loss functions: Theoretical analysis and applications.
\newblock In \emph{International conference on Machine learning}, pages
  23803--23828. PMLR, 2023.

\bibitem[Leng et~al.(2022)Leng, Tan, Liu, Cubuk, Shi, Cheng, and
  Anguelov]{leng2022polyloss}
Zhaoqi Leng, Mingxing Tan, Chenxi Liu, Ekin~Dogus Cubuk, Xiaojie Shi, Shuyang
  Cheng, and Dragomir Anguelov.
\newblock Polyloss: A polynomial expansion perspective of classification loss
  functions.
\newblock \emph{arXiv preprint arXiv:2204.12511}, 2022.

\bibitem[Yang et~al.(2024)Yang, Wang, Gu, and Yang]{yang2024weighted}
Leixin Yang, Haiying Wang, Changgui Gu, and Huijie Yang.
\newblock Weighted signed networks reveal interactions between us foreign
  exchange rates.
\newblock \emph{Entropy}, 26\penalty0 (2):\penalty0 161, 2024.

\bibitem[Barun{\'\i}k et~al.(2017)Barun{\'\i}k, Ko{\v{c}}enda, and
  V{\'a}cha]{barunik2017asymmetric}
Jozef Barun{\'\i}k, Ev{\v{z}}en Ko{\v{c}}enda, and Luk{\'a}{\v{s}} V{\'a}cha.
\newblock Asymmetric volatility connectedness on the forex market.
\newblock \emph{Journal of International Money and Finance}, 77:\penalty0
  39--56, 2017.

\bibitem[Greenwood-Nimmo et~al.(2021)Greenwood-Nimmo, Nguyen, and
  Shin]{greenwood2021measuring}
Matthew Greenwood-Nimmo, Viet~Hoang Nguyen, and Yongcheol Shin.
\newblock Measuring the connectedness of the global economy.
\newblock \emph{International Journal of Forecasting}, 37\penalty0
  (2):\penalty0 899--919, 2021.

\bibitem[Kilic(2017)]{kilic2017contagion}
Erdem Kilic.
\newblock Contagion effects of us dollar and chinese yuan in forward and spot
  foreign exchange markets.
\newblock \emph{Economic Modelling}, 62:\penalty0 51--67, 2017.

\bibitem[Alkan et~al.(2022)Alkan, Alkaya, and Sch{\"u}ller]{alkan2022currency}
Ak{\i}ner Alkan, Ali~Fuat Alkaya, and Peter Sch{\"u}ller.
\newblock Currency exchange rate forecasting with social media sentiment
  analysis.
\newblock In \emph{Intelligent and Fuzzy Techniques for Emerging Conditions and
  Digital Transformation: Proceedings of the INFUS 2021 Conference, held August
  24-26, 2021. Volume 2}, pages 490--497. Springer, 2022.

\bibitem[Ding(2021)]{ding2021conditional}
Liang Ding.
\newblock Conditional correlation between exchange rates and stock prices.
\newblock \emph{The Quarterly Review of Economics and Finance}, 80:\penalty0
  452--463, 2021.

\bibitem[Chantarakasemchit et~al.(2020)Chantarakasemchit, Nuchitprasitchai, and
  Nilsiam]{chantarakasemchit2020forex}
Orawan Chantarakasemchit, Siranee Nuchitprasitchai, and Yuenyong Nilsiam.
\newblock Forex rates prediction on eur/usd with simple moving average
  technique and financial factors.
\newblock In \emph{2020 17th International Conference on Electrical
  Engineering/Electronics, Computer, Telecommunications and Information
  Technology (ECTI-CON)}, pages 771--774. IEEE, 2020.

\bibitem[Gurrib and Elshareif(2016)]{gurrib2016optimizing}
Ikhlaas Gurrib and Elgilani Elshareif.
\newblock Optimizing the performance of the fractal adaptive moving average
  strategy: The case of eur/usd.
\newblock \emph{International Journal of Economics and Finance}, 8\penalty0
  (2):\penalty0 171--179, 2016.

\bibitem[Tornell and Yuan(2012)]{tornell2012speculation}
Aaron Tornell and Chunming Yuan.
\newblock Speculation and hedging in the currency futures markets: Are they
  informative to the spot exchange rates.
\newblock \emph{Journal of Futures Markets}, 32\penalty0 (2):\penalty0
  122--151, 2012.

\bibitem[Ferraro et~al.(2015)Ferraro, Rogoff, and Rossi]{ferraro2015can}
Domenico Ferraro, Kenneth Rogoff, and Barbara Rossi.
\newblock Can oil prices forecast exchange rates? an empirical analysis of the
  relationship between commodity prices and exchange rates.
\newblock \emph{Journal of International Money and Finance}, 54:\penalty0
  116--141, 2015.

\bibitem[Chen and Rogoff(2003)]{chen2003commodity}
Yu-chin Chen and Kenneth Rogoff.
\newblock Commodity currencies.
\newblock \emph{Journal of international Economics}, 60\penalty0 (1):\penalty0
  133--160, 2003.

\bibitem[Lyons(1997)]{lyons1997simultaneous}
Richard~K Lyons.
\newblock A simultaneous trade model of the foreign exchange hot potato.
\newblock \emph{Journal of international Economics}, 42\penalty0
  (3-4):\penalty0 275--298, 1997.

\bibitem[Cashin et~al.(2004)Cashin, C{\'e}spedes, and
  Sahay]{cashin2004commodity}
Paul Cashin, Luis~F C{\'e}spedes, and Ratna Sahay.
\newblock Commodity currencies and the real exchange rate.
\newblock \emph{Journal of Development Economics}, 75\penalty0 (1):\penalty0
  239--268, 2004.

\bibitem[Jadidzadeh and Serletis(2017)]{jadidzadeh2017does}
Ali Jadidzadeh and Apostolos Serletis.
\newblock How does the us natural gas market react to demand and supply shocks
  in the crude oil market?
\newblock \emph{Energy Economics}, 63:\penalty0 66--74, 2017.

\bibitem[Lizardo and Mollick(2010)]{lizardo2010oil}
Radham{\'e}s~A Lizardo and Andr{\'e}~V Mollick.
\newblock Oil price fluctuations and us dollar exchange rates.
\newblock \emph{Energy economics}, 32\penalty0 (2):\penalty0 399--408, 2010.

\bibitem[Basher et~al.(2016)Basher, Haug, and Sadorsky]{basher2016impact}
Syed~Abul Basher, Alfred~A Haug, and Perry Sadorsky.
\newblock The impact of oil shocks on exchange rates: A markov-switching
  approach.
\newblock \emph{Energy Economics}, 54:\penalty0 11--23, 2016.

\bibitem[Pukthuanthong and Roll(2011)]{pukthuanthong2011gold}
Kuntara Pukthuanthong and Richard Roll.
\newblock Gold and the dollar (and the euro, pound, and yen).
\newblock \emph{Journal of Banking \& Finance}, 35\penalty0 (8):\penalty0
  2070--2083, 2011.

\bibitem[Sjaastad(2008)]{sjaastad2008price}
Larry~A Sjaastad.
\newblock The price of gold and the exchange rates: Once again.
\newblock \emph{Resources Policy}, 33\penalty0 (2):\penalty0 118--124, 2008.

\bibitem[Zhang et~al.(2016)Zhang, Dufour, and Galbraith]{zhang2016exchange}
Hui~Jun Zhang, Jean-Marie Dufour, and John~W Galbraith.
\newblock Exchange rates and commodity prices: Measuring causality at multiple
  horizons.
\newblock \emph{Journal of Empirical Finance}, 36:\penalty0 100--120, 2016.

\bibitem[Nazlioglu and Soytas(2012)]{nazlioglu2012oil}
Saban Nazlioglu and Ugur Soytas.
\newblock Oil price, agricultural commodity prices, and the dollar: A panel
  cointegration and causality analysis.
\newblock \emph{Energy Economics}, 34\penalty0 (4):\penalty0 1098--1104, 2012.

\bibitem[Akanni(2020)]{akanni2020returns}
Lateef~Olawale Akanni.
\newblock Returns and volatility spillover between food prices and exchange
  rate in nigeria.
\newblock \emph{Journal of Agribusiness in Developing and Emerging Economies},
  10\penalty0 (3):\penalty0 307--325, 2020.

\bibitem[Nazlioglu et~al.(2013)Nazlioglu, Erdem, and
  Soytas]{nazlioglu2013volatility}
Saban Nazlioglu, Cumhur Erdem, and Ugur Soytas.
\newblock Volatility spillover between oil and agricultural commodity markets.
\newblock \emph{Energy Economics}, 36:\penalty0 658--665, 2013.

\bibitem[Rezitis(2015)]{rezitis2015relationship}
Anthony~N Rezitis.
\newblock The relationship between agricultural commodity prices, crude oil
  prices and us dollar exchange rates: A panel var approach and causality
  analysis.
\newblock \emph{International Review of Applied Economics}, 29\penalty0
  (3):\penalty0 403--434, 2015.

\bibitem[Engel and Wu(2023)]{engel2023forecasting}
Charles Engel and Steve Pak~Yeung Wu.
\newblock Forecasting the us dollar in the 21st century.
\newblock \emph{Journal of International Economics}, 141:\penalty0 103715,
  2023.

\bibitem[Chinn and Meredith(2004)]{chinn2004monetary}
Menzie~D Chinn and Guy Meredith.
\newblock Monetary policy and long-horizon uncovered interest parity.
\newblock \emph{IMF staff papers}, 51\penalty0 (3):\penalty0 409--430, 2004.

\bibitem[Lace et~al.(2015)Lace, Ma{\v{c}}erinskien{\.e}, and
  Bal{\v{c}}i{\=u}nas]{lace2015determining}
Natalja Lace, Irena Ma{\v{c}}erinskien{\.e}, and Andrius Bal{\v{c}}i{\=u}nas.
\newblock Determining the eur/usd exchange rate with us and german government
  bond yields in the post-crisis period.
\newblock \emph{Intellectual Economics}, 9\penalty0 (2):\penalty0 150--155,
  2015.

\bibitem[Afonso and Kazemi(2018)]{afonso2018euro}
Antonio Afonso and Mina Kazemi.
\newblock Euro area sovereign yields and the power of unconventional monetary
  policy.
\newblock \emph{Finance a {\'u}v{\v{e}}r-Czech Journal of Economics and
  Finance}, 68\penalty0 (2):\penalty0 100--119, 2018.

\bibitem[Cecioni(2018)]{cecioni2018ecb}
Martina Cecioni.
\newblock Ecb monetary policy and the euro exchange rate.
\newblock \emph{Bank of Italy Temi di Discussione (Working Paper) No}, 1172,
  2018.

\bibitem[Tonzer(2015)]{tonzer2015cross}
Lena Tonzer.
\newblock Cross-border interbank networks, banking risk and contagion.
\newblock \emph{Journal of Financial Stability}, 18:\penalty0 19--32, 2015.

\bibitem[Ivashina et~al.(2015)Ivashina, Scharfstein, and
  Stein]{ivashina2015dollar}
Victoria Ivashina, David~S Scharfstein, and Jeremy~C Stein.
\newblock Dollar funding and the lending behavior of global banks.
\newblock \emph{The Quarterly Journal of Economics}, 130\penalty0 (3):\penalty0
  1241--1281, 2015.

\bibitem[Dal~Bianco et~al.(2012)Dal~Bianco, Camacho, and Quiros]{dal2012short}
Marcos Dal~Bianco, Maximo Camacho, and Gabriel~Perez Quiros.
\newblock Short-run forecasting of the euro-dollar exchange rate with economic
  fundamentals.
\newblock \emph{Journal of International Money and Finance}, 31\penalty0
  (2):\penalty0 377--396, 2012.

\bibitem[Duffie and Stein(2015)]{duffie2015reforming}
Darrell Duffie and Jeremy~C Stein.
\newblock Reforming libor and other financial market benchmarks.
\newblock \emph{Journal of Economic Perspectives}, 29\penalty0 (2):\penalty0
  191--212, 2015.

\bibitem[Du et~al.(2018)Du, Tepper, and Verdelhan]{du2018deviations}
Wenxin Du, Alexander Tepper, and Adrien Verdelhan.
\newblock Deviations from covered interest rate parity.
\newblock \emph{The Journal of Finance}, 73\penalty0 (3):\penalty0 915--957,
  2018.

\bibitem[Eisenschmidt et~al.(2018)Eisenschmidt, Kedan, and
  Tietz]{eisenschmidt2018measuring}
Jens Eisenschmidt, Danielle Kedan, and Robin~Darius Tietz.
\newblock Measuring fragmentation in the euro area unsecured overnight
  interbank money market.
\newblock \emph{Economic Bulletin Articles}, 5, 2018.

\bibitem[Pan et~al.(2007)Pan, Fok, and Liu]{pan2007dynamic}
Ming-Shiun Pan, Robert Chi-Wing Fok, and Y~Angela Liu.
\newblock Dynamic linkages between exchange rates and stock prices: Evidence
  from east asian markets.
\newblock \emph{International Review of Economics \& Finance}, 16\penalty0
  (4):\penalty0 503--520, 2007.

\bibitem[Tsai et~al.(2019)Tsai, Chen, and Wang]{tsai2019predict}
Yun-Cheng Tsai, Jun-Hao Chen, and Jun-Jie Wang.
\newblock Predict forex trend via convolutional neural networks.
\newblock \emph{Journal of Intelligent Systems}, 29\penalty0 (1):\penalty0
  941--958, 2019.

\bibitem[Pandey et~al.(2018)Pandey, Jagadev, Dehuri, and Cho]{pandey2018review}
Trilok~Nath Pandey, Alok~Kumar Jagadev, Satchidananda Dehuri, and Sung-Bae Cho.
\newblock A review and empirical analysis of neural networks based exchange
  rate prediction.
\newblock \emph{Intelligent Decision Technologies}, 12\penalty0 (4):\penalty0
  423--439, 2018.

\bibitem[Nieh and Lee(2001)]{nieh2001dynamic}
Chien-Chung Nieh and Cheng-Few Lee.
\newblock Dynamic relationship between stock prices and exchange rates for g-7
  countries.
\newblock \emph{The Quarterly Review of Economics and Finance}, 41\penalty0
  (4):\penalty0 477--490, 2001.

\bibitem[Lin(2012)]{lin2012comovement}
Chien-Hsiu Lin.
\newblock The comovement between exchange rates and stock prices in the asian
  emerging markets.
\newblock \emph{International Review of Economics \& Finance}, 22\penalty0
  (1):\penalty0 161--172, 2012.

\bibitem[Phylaktis and Ravazzolo(2005)]{phylaktis2005stock}
Kate Phylaktis and Fabiola Ravazzolo.
\newblock Stock prices and exchange rate dynamics.
\newblock \emph{Journal of international Money and Finance}, 24\penalty0
  (7):\penalty0 1031--1053, 2005.

\bibitem[Inci and Lee(2014)]{inci2014dynamic}
A~Can Inci and Bong~Soo Lee.
\newblock Dynamic relations between stock returns and exchange rate changes.
\newblock \emph{European Financial Management}, 20\penalty0 (1):\penalty0
  71--106, 2014.

\bibitem[Moore and Wang(2014)]{moore2014dynamic}
Tomoe Moore and Ping Wang.
\newblock Dynamic linkage between real exchange rates and stock prices:
  Evidence from developed and emerging asian markets.
\newblock \emph{International Review of Economics \& Finance}, 29:\penalty0
  1--11, 2014.

\bibitem[Tsai(2012)]{tsai2012relationship}
I-Chun Tsai.
\newblock The relationship between stock price index and exchange rate in asian
  markets: A quantile regression approach.
\newblock \emph{Journal of International Financial Markets, Institutions and
  Money}, 22\penalty0 (3):\penalty0 609--621, 2012.

\bibitem[Tah and Ngene(2021)]{tah2021dynamic}
Kenneth~A Tah and Geoffrey Ngene.
\newblock Dynamic linkages between us and eurodollar interest rates: new
  evidence from causality in quantiles.
\newblock \emph{Journal of Economics and Finance}, 45:\penalty0 200--210, 2021.

\bibitem[Agrawal et~al.(2010)Agrawal, Srivastav, and
  Srivastava]{agrawal2010study}
Gaurav Agrawal, Aniruddh~Kumar Srivastav, and Ankita Srivastava.
\newblock A study of exchange rates movement and stock market volatility.
\newblock \emph{International Journal of business and management}, 5\penalty0
  (12):\penalty0 62, 2010.

\bibitem[Andreou et~al.(2013)Andreou, Matsi, and Savvides]{andreou2013stock}
Elena Andreou, Maria Matsi, and Andreas Savvides.
\newblock Stock and foreign exchange market linkages in emerging economies.
\newblock \emph{Journal of International Financial Markets, Institutions and
  Money}, 27:\penalty0 248--268, 2013.

\bibitem[Brunnermeier et~al.(2008)Brunnermeier, Nagel, and
  Pedersen]{brunnermeier2008carry}
Markus~K Brunnermeier, Stefan Nagel, and Lasse~H Pedersen.
\newblock Carry trades and currency crashes.
\newblock \emph{NBER macroeconomics annual}, 23\penalty0 (1):\penalty0
  313--348, 2008.

\bibitem[Cairns et~al.(2007)Cairns, Ho, and McCauley]{cairns2007exchange}
John Cairns, Corrinne Ho, and Robert~N McCauley.
\newblock Exchange rates and global volatility: implications for asia-pacific
  currencies.
\newblock \emph{BIS Quarterly Review, March}, 2007.

\bibitem[Pan et~al.(2019)Pan, Wang, Liu, and Wang]{pan2019improving}
Zhiyuan Pan, Yudong Wang, Li~Liu, and Qing Wang.
\newblock Improving volatility prediction and option valuation using vix
  information: A volatility spillover garch model.
\newblock \emph{Journal of Futures Markets}, 39\penalty0 (6):\penalty0
  744--776, 2019.

\bibitem[Angeletos et~al.(2020)Angeletos, Collard, and
  Dellas]{angeletos2020business}
George-Marios Angeletos, Fabrice Collard, and Harris Dellas.
\newblock Business-cycle anatomy.
\newblock \emph{American Economic Review}, 110\penalty0 (10):\penalty0
  3030--3070, 2020.

\bibitem[Ke et~al.(2019{\natexlab{b}})Ke, Kelly, and Xiu]{ke2019predicting}
Zheng~Tracy Ke, Bryan~T Kelly, and Dacheng Xiu.
\newblock Predicting returns with text data.
\newblock Technical report, National Bureau of Economic Research,
  2019{\natexlab{b}}.

\bibitem[Blei et~al.(2003)Blei, Ng, and Jordan]{blei2003latent}
David~M Blei, Andrew~Y Ng, and Michael~I Jordan.
\newblock Latent dirichlet allocation.
\newblock \emph{Journal of machine Learning research}, 3\penalty0
  (Jan):\penalty0 993--1022, 2003.

\bibitem[Aziz et~al.(2022)Aziz, Dowling, Hammami, and
  Piepenbrink]{aziz2022machine}
Saqib Aziz, Michael Dowling, Helmi Hammami, and Anke Piepenbrink.
\newblock Machine learning in finance: A topic modeling approach.
\newblock \emph{European Financial Management}, 28\penalty0 (3):\penalty0
  744--770, 2022.

\bibitem[MERTER et~al.(2023)MERTER, BALCIO{\u{G}}LU, {\c{C}}EREZ, and
  G{\"o}khan]{merter2023evaluation}
Ar{\c{s}} G{\"o}r Abdullah~K{\"u}r{\c{s}}at MERTER, Yavuz~Selim
  BALCIO{\u{G}}LU, Ar{\c{s}} G{\"o}r~Sedat {\c{C}}EREZ, and {\"O}ZER
  G{\"o}khan.
\newblock Evaluation of annual reports of companies with sentiment analysis: An
  application in bist100 index.
\newblock In \emph{8th International New York Academic Research Congress On
  Humanities and Social Sciences}, pages 197--204, 2023.

\bibitem[Zhou et~al.(2023)Zhou, Kan, Huang, and Silbernagel]{zhou2023guided}
Sulong Zhou, Pengyu Kan, Qunying Huang, and Janet Silbernagel.
\newblock A guided latent dirichlet allocation approach to investigate
  real-time latent topics of twitter data during hurricane laura.
\newblock \emph{Journal of Information Science}, 49\penalty0 (2):\penalty0
  465--479, 2023.

\bibitem[Blei and Lafferty(2006)]{blei2006dynamic}
David~M Blei and John~D Lafferty.
\newblock Dynamic topic models.
\newblock In \emph{Proceedings of the 23rd international conference on Machine
  learning}, pages 113--120, 2006.

\bibitem[Ivanov and Kilian(2005)]{ivanov2005practitioner}
Ventzislav Ivanov and Lutz Kilian.
\newblock A practitioner's guide to lag order selection for var impulse
  response analysis.
\newblock \emph{Studies in Nonlinear Dynamics \& Econometrics}, 9\penalty0 (1),
  2005.

\bibitem[Stock and Watson(2001)]{stock2001vector}
James~H Stock and Mark~W Watson.
\newblock Vector autoregressions.
\newblock \emph{Journal of Economic perspectives}, 15\penalty0 (4):\penalty0
  101--115, 2001.

\bibitem[Musa et~al.(2024)Musa, Ahmad, and Maijamaa]{musa2024application}
Yahaya Musa, Ibrahim Ahmad, and Bilkisu Maijamaa.
\newblock Application of vector autoregression (var) on modelling and
  forecasting average monthly rainfall and temperature.
\newblock \emph{Lloyd Business Review}, pages 1--23, 2024.

\bibitem[Mondal et~al.(2023)Mondal, Srinivasan, and
  Ghosh]{mondal2023multivariate}
Semanto Mondal, Prakash Srinivasan, and Rajib~Chandra Ghosh.
\newblock Multivariate time series forecasting to forecast weight dynamics.
\newblock In \emph{2023 Second International Conference on Advances in
  Computational Intelligence and Communication (ICACIC)}, pages 1--6. IEEE,
  2023.

\bibitem[Schorfheide et~al.(2018)Schorfheide, Song, and
  Yaron]{schorfheide2018identifying}
Frank Schorfheide, Dongho Song, and Amir Yaron.
\newblock Identifying long-run risks: A bayesian mixed-frequency approach.
\newblock \emph{Econometrica}, 86\penalty0 (2):\penalty0 617--654, 2018.

\bibitem[Darst et~al.(2018)Darst, Malecki, and Engelman]{darst2018using}
Burcu~F Darst, Kristen~C Malecki, and Corinne~D Engelman.
\newblock Using recursive feature elimination in random forest to account for
  correlated variables in high dimensional data.
\newblock \emph{BMC genetics}, 19:\penalty0 1--6, 2018.

\bibitem[Zhou et~al.(2016)Zhou, Zhou, and Li]{zhou2016cost}
Qifeng Zhou, Hao Zhou, and Tao Li.
\newblock Cost-sensitive feature selection using random forest: Selecting
  low-cost subsets of informative features.
\newblock \emph{Knowledge-based systems}, 95:\penalty0 1--11, 2016.

\bibitem[Gregorutti et~al.(2017)Gregorutti, Michel, and
  Saint-Pierre]{gregorutti2017correlation}
Baptiste Gregorutti, Bertrand Michel, and Philippe Saint-Pierre.
\newblock Correlation and variable importance in random forests.
\newblock \emph{Statistics and Computing}, 27:\penalty0 659--678, 2017.

\bibitem[Yang and Wang(2022)]{yang2022adaptability}
Mo~Yang and Jing Wang.
\newblock Adaptability of financial time series prediction based on bilstm.
\newblock \emph{Procedia Computer Science}, 199:\penalty0 18--25, 2022.

\bibitem[Ma et~al.(2023)Ma, Mao, Lin, Wu, and Cambria]{ma2023multi}
Yu~Ma, Rui Mao, Qika Lin, Peng Wu, and Erik Cambria.
\newblock Multi-source aggregated classification for stock price movement
  prediction.
\newblock \emph{Information Fusion}, 91:\penalty0 515--528, 2023.

\bibitem[Siami-Namini et~al.(2019)Siami-Namini, Tavakoli, and
  Namin]{siami2019performance}
Sima Siami-Namini, Neda Tavakoli, and Akbar~Siami Namin.
\newblock The performance of lstm and bilstm in forecasting time series.
\newblock In \emph{2019 IEEE International conference on big data (Big Data)},
  pages 3285--3292. IEEE, 2019.

\bibitem[Akiba et~al.(2019)Akiba, Sano, Yanase, Ohta, and
  Koyama]{akiba2019optuna}
Takuya Akiba, Shotaro Sano, Toshihiko Yanase, Takeru Ohta, and Masanori Koyama.
\newblock Optuna: A next-generation hyperparameter optimization framework.
\newblock In \emph{Proceedings of the 25th ACM SIGKDD international conference
  on knowledge discovery \& data mining}, pages 2623--2631, 2019.

\bibitem[Zhou et~al.(2024)Zhou, Dong, and Bao]{zhou2024ship}
Yipeng Zhou, Ze~Dong, and Xiongguan Bao.
\newblock A ship trajectory prediction method based on an optuna--bilstm model.
\newblock \emph{Applied Sciences}, 14\penalty0 (9):\penalty0 3719, 2024.

\bibitem[Lei et~al.(2021)Lei, Zhang, and Song]{lei2021volatility}
Bolin Lei, Boyu Zhang, and Yuping Song.
\newblock Volatility forecasting for high-frequency financial data based on web
  search index and deep learning model.
\newblock \emph{Mathematics}, 9\penalty0 (4):\penalty0 320, 2021.

\bibitem[Diebold and Mariano(2002)]{diebold2002comparing}
Francis~X Diebold and Robert~S Mariano.
\newblock Comparing predictive accuracy.
\newblock \emph{Journal of Business \& economic statistics}, 20\penalty0
  (1):\penalty0 134--144, 2002.

\bibitem[Molodtsova and Papell(2009)]{molodtsova2009out}
Tanya Molodtsova and David~H Papell.
\newblock Out-of-sample exchange rate predictability with taylor rule
  fundamentals.
\newblock \emph{Journal of international economics}, 77\penalty0 (2):\penalty0
  167--180, 2009.

\end{thebibliography}

\end{document}